\documentclass[aps,pra,reprint,twocolumn]{revtex4-2} 
\usepackage{times}
\usepackage{xcolor}
\usepackage{amsmath}
\usepackage{graphicx}
\usepackage{scalerel}
\usepackage{tikz}

\usepackage[unicode,breaklinks]{hyperref}
\hypersetup{
    unicode=true,
    plainpages=false, 
    colorlinks=true,
    linkcolor=blue,
    citecolor=blue,
    filecolor=black,
    urlcolor=blue
}
\usetikzlibrary{svg.path}

\newcommand\bk{\mathbf{k}}
\newcommand\br{\mathbf{r}}
\newcommand\bx{\mathbf{x}}
\newcommand\brho{\boldsymbol{\rho}}
\newcommand\npeak{n_\mathrm{peak}}
\newcommand\gammaQF{\gamma_\mathrm{QF}}
\newcommand\MQ{\mathcal{Q}}
\newcommand\sT{\mathsf{T}}
\newcommand\sR{\mathsf{R}}
\newcommand\Ub{\bar{U}}
\newcommand\elim{\epsilon_{\mathrm{lim}}}
\renewcommand{\Re}{\operatorname{Re}}
\renewcommand{\Im}{\operatorname{Im}}
\definecolor{orcidlogocol}{HTML}{A6CE39}
\tikzset{
  orcidlogo/.pic={
    \fill[orcidlogocol] svg{M256,128c0,70.7-57.3,128-128,128C57.3,256,0,198.7,0,128C0,57.3,57.3,0,128,0C198.7,0,256,57.3,256,128z};
    \fill[white] svg{M86.3,186.2H70.9V79.1h15.4v48.4V186.2z}
                 svg{M108.9,79.1h41.6c39.6,0,57,28.3,57,53.6c0,27.5-21.5,53.6-56.8,53.6h-41.8V79.1z M124.3,172.4h24.5c34.9,0,42.9-26.5,42.9-39.7c0-21.5-13.7-39.7-43.7-39.7h-23.7V172.4z}
                 svg{M88.7,56.8c0,5.5-4.5,10.1-10.1,10.1c-5.6,0-10.1-4.6-10.1-10.1c0-5.6,4.5-10.1,10.1-10.1C84.2,46.7,88.7,51.3,88.7,56.8z};
  }
}

\newcommand{\orcidicon}[1]{\href{https://orcid.org/#1}{\mbox{\scalerel*{
\begin{tikzpicture}[yscale=-1,transform shape]
\pic{orcidlogo};
\end{tikzpicture}
}{|}}}}

\begin{document}
\title{Symmetry and self-bound droplets in dipolar molecular gases}
\author{D.~Baillie\,\orcidicon{0000-0002-8194-7612}}
\affiliation{Dodd-Walls Centre for Photonic and Quantum Technologies, Dunedin 9054, New Zealand}
\affiliation{Department of Physics, University of Otago, Dunedin 9016, New Zealand}
\date{\today}
\begin{abstract}
Recent experiments with degenerate molecular gases dressed by elliptically polarized microwave fields have enabled new control of dipolar interactions via engineered anisotropy. We reveal a symmetry structure of the dipolar interaction that generates degeneracies among the interaction parameters, enabling a classification of spatial symmetries and equilibrium shapes of the gases. Exploiting these symmetries, we analyze solutions including beyond-meanfield quantum fluctuations in the dilute gas regime, and develop a complementary variational theory. We map out the phase diagram of self-bound droplets and characterize their widths, energies, and densities. 
\end{abstract}
\maketitle
\section{Introduction}
Gases of ultracold bialkali molecules have long held the promise of strong long-range and anisotropic dipolar interactions. Early progress with fermions \cite{Ni2008a} was hindered by losses due to chemical reactions \cite{Ospelkaus2010a,Ni2010a}, although deep Fermi degeneracy was ultimately achieved \cite{De-Marco2019a}. By using microwave shielding \cite{Karman2018a,Lassabliere2018a,Anderegg2021a} to limit losses, bosonic molecules were finally condensed in 2024 \cite{Bigagli2023a,Bigagli2024a} (see also \cite{Langen2024a,Lin2023a,Shi2025a}). 

The observation in magnetic atoms of several long-lived localized droplets arising from a trapped dipolar condensate \cite{Kadau2016a} and macrodroplets \cite{Chomaz2016a} led to the realization of a self-bound dipolar droplet \cite{Schmitt2016a}, stable in free-space without external confinement, stabilized by quantum fluctuations \cite{Petrov2015a} (see also \cite{Lahaye2009a,Chomaz2023a,Chomaz2026a,Baillie2016b,Bisset2016a,Wachtler2016a,Wachtler2016b}). Self-bound droplets in two-component atomic condensates have since been observed \cite{Cabrera2018a,Semeghini2018a}. 

Experiments now bring this physics to molecular gases with self-bound droplets recently reported \cite{Zhang2026a}, which demonstrated several important advances. First, higher dipolar interaction strengths than can be achieved with atomic magnetic gases. Second, the sign of the effective dipolar interaction can be changed for molecular gases. The change of dipolar interaction sign was proposed for \emph{anti-dipolar} atomic gases by fast rotation of the magnetic field \cite{Giovanazzi2002a,Baillie2020a}, and modification of the interaction by rotation was realized experimentally for trapped atoms \cite{Tang2018a}. Theory showed that such trapped gases prefer a pancake shape, and with confinement perpendicular to the dipole orientation can result in a stack of `pancakelets'  \cite{Wenzel2018c} (see also \cite{Mukherjee2023b,Mukherjee2023c}). However, lifetimes of atomic gases with rapidly rotating magnetic fields have been limited, whereas molecular gases show impressive lifetimes \cite{Zhang2026a}. Finally, with interactions controlled by elliptically polarized microwave fields demonstrated in \cite{Zhang2026a}, interactions of molecular gases have increased anisotropy beyond the cylindrical symmetry of the atomic dipolar interaction. 

Existing theory for molecular Bose gases has investigated molecular condensates \cite{Jin2025a}, droplets \cite{Zhang2025c,Langen2025a,Ciardi2025a} and layering \cite{Polterauer2025a}, and has been limited to the cylindrically symmetric interaction; only very recent work considers elliptically controlled anisotropy in the context of supersolids \cite{Cardinale2025a,Zhang2025a}.

In this work, we uncover a symmetry framework for the microwave-dressed dipolar interaction, showing that each combination of interaction parameters belongs to a sextet of degenerate configurations related by coordinate permutations. This structure organizes the interaction landscape, revealing where cylindrical symmetry is preserved or broken, and classifying the equilibrium shapes of the cloud.  This symmetry analysis follows directly from the form of the interaction, and is therefore not specific to the later theoretical treatment. We then apply the symmetries to the extended Gross-Pitaevskii equation (eGPE) including beyond meanfield effects, and we develop variational solutions that capture the key anisotropy effects. The eGPE approach is not valid when $a_s<0$ or in the strongly interacting regime. Focusing on self-bound droplets, we derive the phase diagram, first in the thermodynamic limit, and then for finite molecule number. We demonstrate the properties of self-bound droplets and compare the energy and peak density to the thermodynamic limit scaling. 

\section{Interaction}%
We consider a system with effective inter-molecular dipole-dipole potential $U(\br)$, and $s$-wave interactions with interaction strength $g_s=4\pi\hbar^2a_s/m$ for molecules with $s$-wave scattering length $a_s$. We assume the molecules are shielded by a microwave field \cite{Deng2023a,Karman2025a,Zhang2025a,Deng2025a,Schindewolf2026a} with ellipticity $\xi$, giving 
\footnote{Using $a_{\mathrm{d}m}$ defined in \cite{Zhang2026a}, our $\epsilon_m = a_{\mathrm{d}m}/3a_s$ for $m=0$, $2$. Using $\epsilon_{3,2}$ defined in \cite{Zhang2025a}, our $\epsilon_2=\epsilon_{3,2}/\sqrt3$.}

\begin{align}
    U(\br) = \frac{3|g_s|}{4\pi r^3}[\epsilon_0(1-3\cos^2\theta) + \sqrt3\epsilon_2\sin^2\theta\cos2\phi],\label{e:Ur}
\end{align}
where $\theta$ is the angle between $\br$ and the $z$ axis, and $\phi$ is the angle between $\brho = (x,y)$ and the $x$ axis. 
The first dipolar interaction term is the usual $\phi$ independent term $Y_2^0$ with relative dipole interaction strength $\epsilon_0$ \footnote{Not to be confused with the vacuum permittivity.}, and the second $Y_2^2+Y_2^{-2}$ term depends on both $\theta$ and $\phi$, with relative dipole interaction strength $\epsilon_2\propto\sin2\xi$. The contribution from $Y_2^{\pm1}$ is generally small \cite{Karman2025a}. As a wide range of positive and negative dipolar interactions is possible experimentally \cite{Yuan2025a,Zhang2026a,Karman2025a}, we allow for positive, zero, and negative values of $\epsilon_0$ and $\epsilon_2$.
With no ellipticity of the microwave field, $\xi=0$ so that $\epsilon_2=0$, results are equivalent to those for dipolar atoms when $\epsilon_0>0$ and anti-dipolar atoms when $\epsilon_0<0$.

\emph{Symmetry--}
\begin{figure}
    {\includegraphics[trim=3 3 5.5 0,clip=true,width=0.9\linewidth]{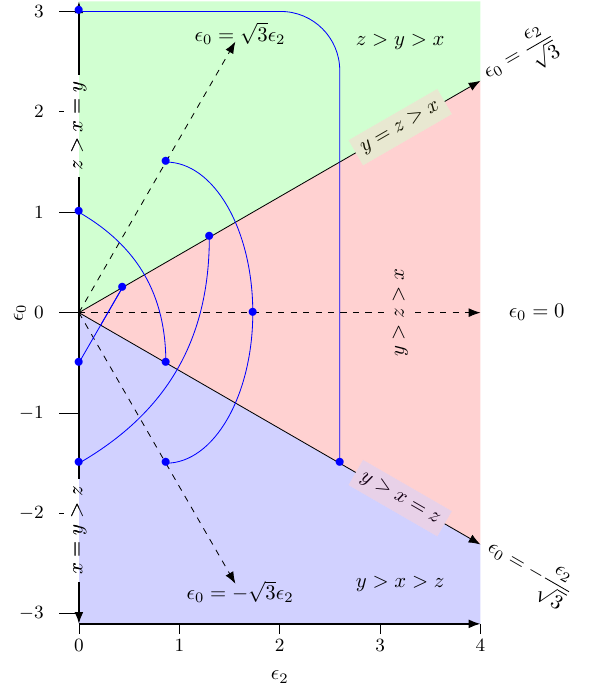}}% 
    \caption{\label{f:sym}Symmetry of dipolar molecular gases and relative cloud widths. The blue dots show equivalent states under permutations of the coordinates (the shape of the connecting blue lines is arbitrary). There is also simple reflection symmetry about the line $\epsilon_2=0$: the $\epsilon_2<0$ results and their connection to the dots are not shown (if three dots are linked above, there are another three for $\epsilon_2<0$, and if two are shown there is another one for $\epsilon_2<0$). The three colored regions (six including $\epsilon_2<0$) demarcate domains with different relative widths. The notation $x=y>z$ means that the cloud has $x,y$ symmetry, and the $x,y$ width is greater than the $z$ width, unless broken by a confining potential. The dashed lines indicate $\epsilon_0=0$ and other members of its sextet.}     
\end{figure}
The dependence of results on $(\epsilon_0,\epsilon_2)$ is via functions of $U(\br)$ which can be written
\begin{align}
    U(\br) = \frac{3|g_s|}{4\pi r^5}[(\epsilon_0+\sqrt3\epsilon_2)x^2 +(\epsilon_0-\sqrt3\epsilon_2)y^2 - 2\epsilon_0z^2],\label{e:Ursym}
\end{align}
so the ground state and its energy are unchanged by those changes to $\epsilon_0$ and $\epsilon_2$ that correspond to permutations of the coordinates $\{x,y,z\}$, for example states with $\epsilon_0$, $\epsilon_2$ are equivalent to those with $\epsilon_0$, $-\epsilon_2$, exchanging the roles of $x$ and $y$. 

That is, each state belongs to a sextet with the same energy and wavefunction related by coordinate permutations. The transformation between the parameter sets is given by $\boldsymbol{\epsilon}' =\sT\boldsymbol{\epsilon}$ where  $\boldsymbol{\epsilon} = (\epsilon_0,\epsilon_2)^T$ and $\sT$ is any product of the reflections $\sT\!_n = \begin{pmatrix} \cos\vartheta & \sin\vartheta\\ \sin\vartheta & -\cos\vartheta \end{pmatrix}$ and rotations $\sR_n = \begin{pmatrix} \cos\vartheta & -\sin\vartheta\\ \sin\vartheta & \cos\vartheta \end{pmatrix}$ in parameter space, for $\vartheta=2\pi n/3$ with $n=0,1,2$, i.e. $\sT$ is from the dihedral group $D_3$. 

Eq.~\eqref{e:Ursym} also shows when the system has cylindrical symmetry, assuming the trap, if any, has the same symmetry: when $\epsilon_2=0$ they have $x,y$ symmetry, when $\epsilon_0=\epsilon_2/\sqrt3$, they have $y,z$ symmetry, and when $\epsilon_0=-\epsilon_2/\sqrt3$ they have $x,z$ symmetry, i.e. the symmetry axes of the $D_3$ group. On a symmetry axis, the sextet collapses to three distinct states due to degeneracy.

This also shows that near the $\epsilon_0>0$ axis and near $\epsilon_0=-\epsilon_2/\sqrt3$ the gas is prolate, whereas near the $\epsilon_0<0$ axis and near $\epsilon_0=\epsilon_2/\sqrt3$ the cloud is oblate, subject to any modification by a trap.

Finally, Eq.~\eqref{e:Ursym} establishes the relative widths of the gas, up to potential modifications by a trap. For example, when $\epsilon_0>\epsilon_2/\sqrt3>0$ the coefficients are in decreasing order of $(x,y,z)$, so the real-space widths increase in that order to minimize energy.  

In what follows, we particularly focus on three cases: 
prolate symmetry $(\epsilon_0>0,\:\epsilon_2=0)$, 
oblate  symmetry $(\epsilon_0<0,\:\epsilon_2=0)$, 
and asymmetric, midway between the symmetry axes $(\epsilon_0=0)$.
After transformations, these three cases cover twelve lines, every 30\textdegree, in parameter space. 

The above discussion on the symmetry and lengths is summarized in Fig.~\ref{f:sym}. The blue dots, linked by blue lines, are equivalent states. As we omit $\epsilon_2<0$, states are in groups of two if on a symmetry axis, otherwise groups of three. The $D_3$ symmetries tile the plane: each of the three colored regions has all of the different possible states, subject to permutation of the coordinates. 

\section{\protect\lowercase{e}GPE}%
As an application, we consider the ground state wavefunction, $\psi(\bx)$, given by the solution to the time-independent eGPE
\begin{align}
    \left[-\frac{\hbar^2 \nabla^2}{2m} + V(\bx) + \Phi(\bx) + \gammaQF|\psi|^3\right]\psi = \mu\psi, \label{e:GPE}
\end{align}
where $\mu$ is the chemical potential, $V(\bx)$ is the trap, and $\psi$ is normalized to the total number of molecules $\int d\bx\,n(\bx)=N$, with density $n(\bx) = |\psi(\bx)|^2$. 

The effective two-body interaction potential is $\Phi(\bx) = g_s|\psi(\bx)|^2 + \int d\bx'\,U(\bx-\bx')|\psi(\bx')|^2$.  As $\epsilon_0$ or $\epsilon_2$ can be zero, the dipole lengths $\epsilon_0 a_s$ and $\epsilon_2 a_s$ are not useful lengths for all possible parameters, and we use the $s$-wave scattering length, $a_s$, as our length unit, which we take to be positive, and $E_0=\hbar^2/ma_s^2$ as our energy unit. The interaction in these units is then completely parameterized by $\epsilon_0$ and $\epsilon_2$. 

Quantum fluctuations are included in the local density approximation with $\gammaQF = 32g_s a_s^{3/2}\MQ_5(\epsilon_0,\epsilon_2)/3\sqrt\pi$, where 
\begin{align}
    \MQ_5(\epsilon_0,\epsilon_2) &=\int \frac{d\Omega_k}{4\pi}\, [1+\Ub(\bk)]^{5/2},\label{e:Q5}
\end{align}
and $g_s\Ub(\bk)$ is the Fourier transform of $U(\br)$ [see Eq.~\eqref{e:UbR}].
There is an attractive component to the interaction potential, i.e. $\Ub(\bk)<-1$, when
\begin{align}
    \epsilon_0+\sqrt3|\epsilon_2|>1 \text{ or }\epsilon_0<-\tfrac12, \label{e:tlbound}
\end{align}
and then $\MQ_5$ has an imaginary part due to instabilities of the homogeneous dipolar gas, which we discard \cite{Wachtler2016a,Bisset2016a}; we show the extent of the imaginary part in our results [see also \eqref{e:alphared}].  For certain parameters $\MQ_5$ can be evaluated, including $\MQ_5(\epsilon_0,0)$ for $\epsilon_0>0$ \cite{Lima2011a,Lima2012a}, $\Re\{\MQ_5(\epsilon_0,0)\} = 5\pi (1 - \epsilon_0)^3/(32 \sqrt{3|\epsilon_0|})$ for $\epsilon_0 < -\frac12$, and $\MQ_5(0,\epsilon_2)$ given by \eqref{e:Q5special}. Otherwise, we integrate over one angle numerically using \eqref{e:absQ5}.

The eGPE depends on $\epsilon_0$ and $\epsilon_2$, both in $\Phi(\bx)$ and in $\MQ_5$, only via $U(\br)$, so the symmetries of the previous section apply, subject to the trap. For example, the wavefunction with $(\epsilon_0<0,\epsilon_2=0)$ and trap $V(\bx) = \frac12m[\omega_\rho^2(x^2+y^2)+\omega_z^2z^2]$, is equal to that with $(\epsilon'_0=|\epsilon_0|/2,\epsilon'_2=\sqrt3|\epsilon_0|/2)$ and $\omega'_y=\omega'_z=\omega_\rho, \omega'_x=\omega_z$ after permutation of coordinates.

In the following we will use the energy to compare solutions, which is given by
\begin{align}
  E &= \int d\bx\,\psi^*\Bigl[-\frac{\hbar^2\nabla^2}{2m} + V(\bx) + \frac12\Phi(\bx) + \frac25\gammaQF|\psi|^3\Bigr]\psi.\label{e:E}
\end{align}

\section{Variational}%
We consider a large class of ans\"atze where the density $n(\bx)$ depends on $\bx$ in the scaling form $n(\bx)=n_s(s)$ for some function $n_s$, with $s^2=x^2/l_x^2+y^2/l_y^2+z^2/l_z^2$. The variational Gaussian we consider below is an example, as is Thomas-Fermi. 
Then
$\int d\bx\,\Phi(\bx)n(\bx) = g_s(1-\epsilon_0 f - \sqrt3\epsilon_2 f_2)\int\! d\bx\,[n(\bx)]^2$ \footnote{By taking the Fourier transform, changing variables to $u_i=l_ik_i$, integrating the angular variables, and changing variables back to $k_i$.},
where the functions $f(l_x/l_z,l_y/l_z)$ and $f_2(l_x/l_z,l_y/l_z)$ are defined in Appendix~\ref{s:anisofun}. 
The function $f(x,y)$ is given in terms of incomplete elliptic integrals in \cite{Giovanazzi2006a,Glaum2007b}\footnote{We align our dipoles along $z$ whereas \cite{Giovanazzi2006a} aligns along $x$, but our function $f(x,y)$ is exactly the same as theirs.}. After the transformations $\boldsymbol{\epsilon}' =\sT\boldsymbol{\epsilon}$,  $\epsilon_0 f(l_x/l_z,l_y/l_z) + \sqrt3\epsilon_2 f_2(l_x/l_z,l_y/l_z) = \epsilon'_0 f(l'_x/l'_z,l'_y/l'_z) + \sqrt3\epsilon'_2 f_2(l'_x/l'_z,l'_y/l'_z)$ where $l'_i$ is the transformed length. For example, after $\sT_0$ which exchanges $x$ and $y$, $l'_x/l'_z=l_y/l_z$ which gives us the relations $f(y,x)=f(x,y)$ and $f_2(y,x)=-f_2(x,y)$, so $f_2(x,x)=0$. After $\sT_1$ we usefully get the functional form of $f_2$ in terms of the well-known $f$
\begin{align}
    f_2(x,y) = \frac13[f(1/y,x/y)-f(1/x,y/x)], 
\end{align}
with the limits $f_2(x,\infty)=1/(1+x)$ and $f_2(x,0)=-1$ for $x\ne0$ \footnote{For $0<y\ll1$, as $x$ is reduced $f_2(x,y)$ changes sharply from $\approx-1$ for $x\gg y$, through zero at $x=y$, to $1$ at $x=0$.}. 
The range is $-1\le f_2(x,y)\le 1$, and 
$f_2(x,1) = f(x,1) = -\frac12f_s(1/x)$, with $f_s(x)\equiv f(x,x)$ \cite{Santos2000a}.

In the following we specialize to a variational Gaussian ansatz 
    $\psi_v(\bx) = \sqrt{\frac{N}{\pi^{3/2}l_xl_yl_z}} e^{-\frac12(x^2/l_x^2+y^2/l_y^2+z^2/l_z^2)}$. 

\section{Results}%
We consider self-bound droplets of dipolar molecules, i.e. $V(\bx)=0$. The domain of self-bound droplets in the thermodynamic limit, $N\to\infty$, is given by \eqref{e:tlbound}. 
The formation of self-bound droplets also requires a sufficient number of molecules. The phase boundary between self-bound droplets and the dispersed gas is found by varying $N$, $\epsilon_0$, $\epsilon_2$ so that $E=0$, using stationary states which solve Eq.~\eqref{e:GPE}. 

\begin{figure}
    {\includegraphics[trim=1 35 10 34,clip=true,width=\linewidth]{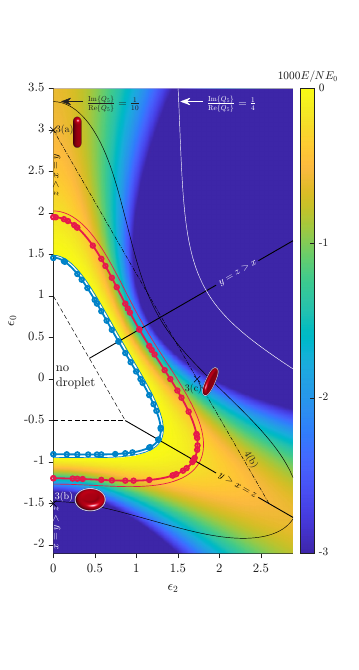}}%
    \caption{\label{f:pdfixedN}Phase diagram for droplets of dipolar molecular gases, with no droplet in the white region near the origin, and stable self-bound droplets outside this region. The energetic stability boundary, $E=0$, is for $N=1000$ using the eGPE (red circles and thick curve) and variational (thin red curve), $N=5000$ using the eGPE (blue circles and thick curve) and variational (thin blue curve), and the thermodynamic limit, Eq.~\eqref{e:tlbound} (dashed lines). Also shown is the variational energy for $N=5000$ (background color), contours where $\Im\{\MQ_5\} = \Re\{\MQ_5\}/10$ (thin black curve) and $\Im\{\MQ_5\} = \Re\{\MQ_5\}/4$ (thin white curve) [see also \eqref{e:alphared}], and the symmetry axes from Fig.~\ref{f:sym} (black lines at $\epsilon_0=\pm\epsilon_2/\sqrt3$). The labels on the symmetry axes show the relative lengths from Fig.~\ref{f:sym}. The cases from Fig.~\ref{f:profile} are marked with $\times$ and the case from Fig.~\ref{f:props}(b) is marked with a dash-dotted line.}
\end{figure}
The phase diagram  for $N=1000$ (red) and $N=5000$ (blue) is given in Fig.~\ref{f:pdfixedN}, showing that the boundary approaches the thermodynamic limit \eqref{e:tlbound} as $N$ increases. The variational energies are also shown (background color). The boundary from eGPE calculations (circles and thick curve) extends to slightly closer to the thermodynamic limit than the variational Gaussian (thin curve), as the eGPE gives lower energy states.  

Compared to the $(\epsilon_0>0,\epsilon_2=0)$ case, the lower energies for the other two cases, $(\epsilon_0<0,\:\epsilon_2=0)$ 
and $\epsilon_0=0$, are striking. To understand this, we consider the large $N$ limit (see Appendix~\ref{s:largeN}),
\begin{align}
    \frac{E}{NE_0} &= -\frac{50\pi^2(\elim-1)^3}{3[64\MQ_5(\epsilon_0,\epsilon_2)]^2}, \label{e:Escale}
\end{align}
where $\elim  = -\min_{\bk}\Ub(\bk)$, i.e.
\begin{align}
    \elim  &= \begin{cases}
      \epsilon_0+\sqrt3|\epsilon_2|, & \epsilon_0\ge-|\epsilon_2|/\sqrt3, \\
    2|\epsilon_0|, & \epsilon_0\le-|\epsilon_2|/\sqrt3.
    \end{cases} \label{e:eeffbounds}
\end{align}
The energy \eqref{e:Escale} is achieved with peak density (see also \cite{Ferrier-Barbut2016a,Dalibard2024a})
\begin{align}
    \npeak a_s^3 &= \frac{25\pi(\elim-1)^2}{[64\MQ_5(\epsilon_0,\epsilon_2)]^2}, \label{e:nscale} 
\end{align}
i.e. independent of $N$ like a liquid. The three cases marked in Fig.~\ref{f:pdfixedN}, all have $\elim=3$, but $[\MQ_5(3,0)]^2\approx10[\MQ_5(0,\sqrt3)]^2\approx20[\MQ_5(-1.5,0)]^2$, which explains the bulk of the relative energy differences. 

As expected from the symmetry, ascending near $\epsilon_0=\epsilon_2/\sqrt3$ gives rapidly decreasing energy, equivalent to descending near the $\epsilon_0<0$ axis, whereas descending near $\epsilon_0=-\epsilon_2/\sqrt3$ the decrease in energy is slower, equivalent to the $\epsilon_0>0$ axis.

Two contours of the imaginary part of the quantum fluctuation term are shown (note that these are independent of $N$), revealing an appreciable region where droplets are formed, before the imaginary part becomes significant.   

\begin{figure}
    {\includegraphics[trim=0 14 0 10,clip=true,width=\linewidth]{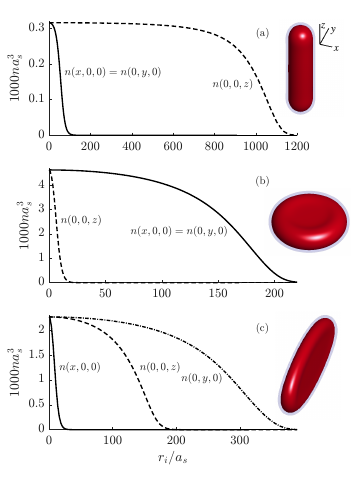}}% 
    \caption{\label{f:profile} Profiles for $N=5000$ and 
    (a) $\epsilon_0=3$, $\epsilon_2=0$, 
    (b) $\epsilon_0=-1.5$, $\epsilon_2=0$ 
    (c) $\epsilon_0=0$, $\epsilon_2=\sqrt3$ for the $x$ axis (solid), $y$ axis (dash-dotted), and $z$ axis (dashed). On the right (not to scale) are isodensity surfaces at $90\%$ (red) and $5\%$ (blue) of peak density.}
\end{figure}
Example profiles of droplets are given in Fig.~\ref{f:profile}. The case $\epsilon_0=3$, $\epsilon_2=0$ is shown in Fig.~\ref{f:profile}(a) with a prolate profile, and the characteristic liquid-like plateau in density in the bulk along $z$. The case $\epsilon_0=-1.5$, $\epsilon_2=0$ is shown in Fig.~\ref{f:profile}(b) with an oblate profile. Fig.~\ref{f:profile}(c) shows the profile on the $\epsilon_2$ axis for $\epsilon_2=\sqrt3$ showing distinct profiles on the three axes. 

For $\epsilon_0=0$, we minimize the variational energy to find that the ratio $\lambda\equiv l_y/l_z$ is constant in the limit $l_x\ll l_z$, given by $(\lambda^2-5 \lambda^4) K(1-\lambda^2)=(2-5\lambda^2-\lambda^4) E(1-\lambda^2)$, with $E$ and $K$ the complete elliptic integrals, i.e., $\lambda\approx1.964$. We also find that all our eGPE results for $\epsilon_0=0$, which we have computed for $1.25\le\epsilon_2\le6$ and $10^2\le N\le10^5$, have $l_y/l_z\in 2\pm 0.1$.

\begin{figure}
    {\includegraphics[trim=5 0 15 0,clip=true,width=\linewidth]{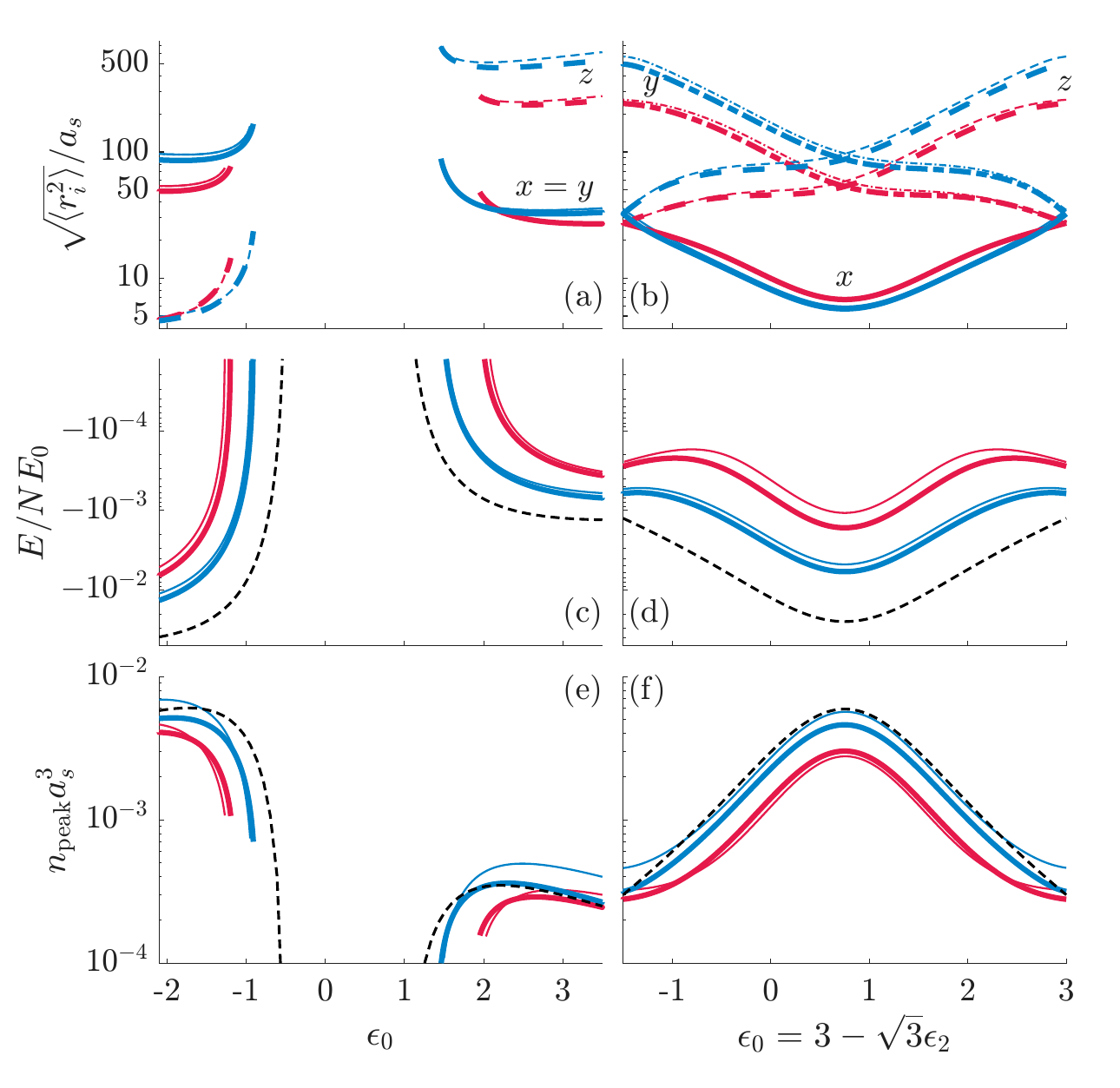}}%
    \caption{\label{f:props}Properties of droplets for $N=1000$ (red) and $N=5000$ (blue), thick curves are eGPE and thin curves are variational. (a,b) RMS widths along the $x$ (solid), $y$ (dash-dotted), and $z$ (dashed) axes [in (a) $\sqrt{\langle y^2\rangle}$ is equal to and obscured by $\sqrt{\langle x^2\rangle}$; for variational $\sqrt{\langle r_i^2\rangle}=l_i/\sqrt2$]. (c,d) Energy per particle, also showing Eq.~\eqref{e:Escale} (black dashed). (e,f) Peak density, also showing Eq.~\eqref{e:nscale} (black dashed).  In (a,c,e) results are for $\epsilon_2=0$, and (b,d,f) for $\epsilon_0+\sqrt3\epsilon_2=3$.}
\end{figure}
Properties of the droplets are shown in Fig.~\ref{f:props}. The long width(s)  increase significantly from $N=1000$ (red) to $N=5000$ (blue), whereas the shortest width(s) show a smaller increase and sometimes decrease. Fig.~\ref{f:props}(b) shows the widths starting from a cylindrically symmetric ($x=y$) prolate droplet on the left at $\epsilon_0=3$, $\epsilon_2=0$ progressing to a cylindrically symmetric ($y=z$) oblate droplet in the middle at $\epsilon_0=3/4$, $\epsilon_2=3\sqrt3/4$, and ending on a cylindrically symmetric ($x=z$) prolate droplet on the right at $\epsilon_0=-3/2$, $\epsilon_2=3\sqrt3/2$. The minimum width in our results is the axial width of the oblate droplet for $\epsilon_0\approx-2$ with $\sqrt{\langle z^2\rangle}\approx 5a_s$. The possibility of a self-bound droplet entering a crossover two-dimensional regime is of interest.

Fig.~\ref{f:props}(c) shows the energy per particle, which decreases sharply from $E=0$ at the phase boundary before quantum fluctuations dominate, i.e. $\MQ_5$ dominates over $\elim$ in~\eqref{e:Escale}. In Figs.~\ref{f:props}(b,d,f), $\elim=3$ is constant, so the variation in Eqs.~\eqref{e:Escale}, \eqref{e:nscale} is entirely due to quantum fluctuations.
Fig.~\ref{f:props}(e) shows the peak density increase steeply as $|\epsilon_0|$ increases away from the critical point. However for larger $|\epsilon_0|\gtrsim2$ the peak density decreases, due again to quantum fluctuations. 
In Fig.~\ref{f:props}(f) the peak density is largest for a cylindrically symmetric oblate droplet.
Figs.~\ref{f:props}(e,f) show that our results are in the dilute gas regime as $n_\mathrm{peak}|a_s\epsilon_i|^3\ll1$.

\section{Conclusions}
We analyzed the symmetry landscape of degenerate molecular gases dressed by elliptically polarized microwave fields, and showed that states form sextets under $D_3$ symmetry that tile the interaction plane, representing the same state with permuted coordinates, capturing the degeneracies of the dipolar interaction landscape. We identified where the system has cylindrical symmetry, and when such a gas is prolate or oblate, and partitioned parameter space by the ordering of the principal widths of the cloud. 
We identified three axes in parameter space (prolate, oblate, and asymmetric), which together represent every 30\textdegree{} in parameter space, due to the symmetry.
We used the symmetries to classify the solutions of the eGPE, and to derive the analytic variational interaction energy. 
Then focusing on self-bound droplets with $a_s>0$ and in the dilute gas regime, we identified the phase diagram in the thermodynamic limit, and computed the finite $N$ phase boundary. We found a limiting form of the energy and peak density of a self-bound droplet, and compared it to numerical results at finite number. We found the asymptotic aspect ratio of the two long widths of our key asymmetric case, $\epsilon_0=0$.
We illustrated the profiles of self-bound droplets and their widths, energy, and peak density. 

The tiling of the interaction plane invites direct experimental verification by scanning $(\epsilon_0,\epsilon_2)$, relating self-bound profiles at different elements of the sextet, and/or by interchanging trap frequencies relative to the polarization direction for trapped gases. Further work will investigate the excitation spectra \cite{Baillie2017a} of self-bound droplets of molecules. 

\begin{acknowledgments}
    We acknowledge useful discussions with Blair Blakie, use of NZ eScience Infrastructure (NeSI) high-performance computing facilities, and support from the Marsden Fund of the Royal Society of New Zealand.
\end{acknowledgments}

\section*{Data availability}The data that support the findings of this article are openly available \cite{Baillie2026b}.

\appendix
\renewcommand\theequation{\Alph{section}\arabic{equation}}
\section{Momentum-space interaction\label{s:momint}}%
We use a grid that is shaped to amply cover our density, which is generally a different size in each direction, and transform onto a zero-padded grid in momentum space to use a spherical cutoff for the dipolar interaction. The cutoff Fourier transform is found using the spherical wave expansion of the plane wave \cite{JacksonBook1999a} 
\begin{align}
    &\int_{r<R} d\br\,e^{-i\bk\cdot\br} \frac3{4\pi r^3}\sum_{m=-2}^2c_mY_2^m(\theta,\phi), \notag\\
    &= -3\sum_{m=-2}^2 c_m Y_2^{m*}(\theta_k,\phi_k)\int_0^Rdr\,r^{-1}j_2(kr),\\
    &= -s(kR)\sum_{m=-2}^2 c_m Y_2^{m*}(\theta_k,\phi_k),
\end{align}
where $s(\kappa) = 1 + 3\kappa^{-2}\cos\kappa-3\kappa^{-3}\sin\kappa$, 
$\theta_k$ is the angle between $\bk$ and $k_z$, and $\phi_k$ is the angle between $(k_x,k_y)$ and the $k_x$ axis. Using the $c_m$ for our interaction gives \cite{Ronen2006a,Cardinale2025a}
\begin{align}
    \frac{\Ub^R(\bk)}{s(kR)}=\epsilon_0(3\cos^2\theta_k-1)-\sqrt3\epsilon_2\sin^2\theta_k\cos 2\phi_k. \label{e:UbR}
\end{align}
For $R\to\infty$, $s(kR)\to 1$, and $\Ub^R(\bk) \to \Ub(\bk)$.

\section{Quantum fluctuations coefficient \label{s:LHY}}%
The function $\Ub(\bk)$ is real, so for a given $\bk$, the principal root $[1+\Ub(\bk)]^{5/2}$ is positive real or positive imaginary, and $\MQ_5(\epsilon_0,\epsilon_2)$ is complex with positive real and imaginary parts if in the region \eqref{e:tlbound}, and real otherwise. First we consider $\epsilon_2=0$
\begin{align}
\MQ_5(x,0) &=  \frac{5(x-1)^3}{16\sqrt{3x}}\Bigl[\ln(\sqrt{1+2x}- \sqrt{3x})-\frac{\ln(1-x)}2\Bigr]\notag\\
&+ \frac1{16}(11+4x+9x^2)\sqrt{1+2x}. \label{e:MQ5x0}
\end{align}
\begin{figure}
    \includegraphics[trim=21 15 53 23,clip=true,width=\linewidth]{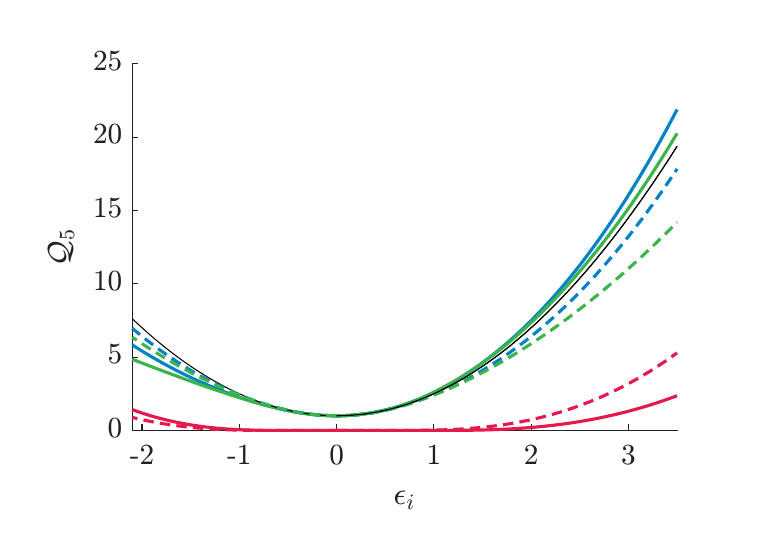}
    \caption{\label{f:Q5}Quantum fluctuations coefficient along $\epsilon_2=0$ (solid) and $\epsilon_0=0$ (dashed). Shown are $\Re\{\MQ_5\}$ (blue), $\Im\{\MQ_5\}$ (red), $\MQ'_5$ \eqref{e:alphared} (green), and the small parameter approximation \eqref{e:Q5small} (thin black).}
\end{figure}

For $\epsilon_2\ne0$, we pull out a $\phi_k$ dependent factor that is imaginary for some parameters, before integrating over $\theta_k$, so we must take the absolute value of the real part before numerically integrating over $\phi_k$ 
\begin{align}
    \Re\{\MQ_5(\epsilon_0,\epsilon_2)\} &= \frac2\pi\int_0^{\pi/2}d\phi_k \left|\Re\{I(\phi_k)\}\right|, \label{e:absQ5} \\
    \Im\{\MQ_5(\epsilon_0,\epsilon_2)\} &= \frac2\pi\int_0^{\pi/2}d\phi_k \Im\{I(\phi_k)\},  \\
    I(\phi_k) = (1-\tfrac2{\sqrt3}&\epsilon_2\cos2\phi_k)^{5/2}\MQ_5\Bigl(\frac{\sqrt3\epsilon_0+\epsilon_2\cos2\phi_k}{\sqrt3-2\epsilon_2\cos2\phi_k},0\Bigr). 
\end{align}
The integral can be evaluated for these two cases
\begin{align}
    &\MQ_5(\epsilon_0,\epsilon_2)\notag\\
    &=\begin{cases}
        {}_3F_2(-\frac54,-\frac34,\frac12;\frac34,\frac54;3\epsilon_2^2) & \epsilon_0=0\\
\frac{5\pi}{32} (\frac32 + \sqrt3 |\epsilon_2|)^{5/2} {}_2F_1\bigl(-\frac52, \frac12; 1; \frac{4 |\epsilon_2|}{2 |\epsilon_2|+\sqrt3}\bigr) & \epsilon_0=-\frac12
\end{cases}\!\!. \label{e:Q5special}
\end{align}
The real part of $\mathcal{Q}_5$ retains the real terms in the integrand of Eq.~(A2) of Ref.~\cite{Bisset2016a} even where the Bogoliubov energy term is imaginary. An alternative is to exclude the entire contribution of unstable modes to that integral \cite{Saito2016a}, i.e., to use Cutoff III of Ref.~\cite{Bisset2016a} generalized to include $\epsilon_2$ (evaluated using Eq.~(A10) of Ref.~\cite{Bisset2016a}, noting \footnote{The prefactor of Eq.~(A10) of \cite{Bisset2016a} should be $1/128$, not $1/64$.}). This defines $\mathcal{Q}'_5$, which we evaluate analytically. With $\alpha\equiv\Im\{\MQ_5\}/\Re\{\MQ_5\}$,
\begin{align}
    \frac{\MQ'_5(\epsilon_0,\epsilon_2)}{\Re\{\MQ_5(\epsilon_0,\epsilon_2)\}} &= 1 - \frac{11}{16}\alpha
    \approx \begin{cases}
        0.93 & \alpha= \frac1{10},\\
        0.83 & \alpha = \frac14.
    \end{cases}\label{e:alphared}
\end{align}
The results for $\MQ_5$ along the axes are shown in Fig.~\ref{f:Q5}, along with the small $\epsilon_0$ and $\epsilon_2$ limit \footnote{For larger parameters $\MQ_5$ increases like $|\epsilon_0|^{5/2}$ or $|\epsilon_2|^{5/2}$ [see Eq.~\eqref{e:Q5}], but the imaginary part of $\MQ_5$ will then be significant.}
\begin{align}
    \MQ_5(\epsilon_0,\epsilon_2) &\approx 1 + \frac32(\epsilon_0^2 + \epsilon_2^2).\label{e:Q5small}
\end{align}

\section{Anisotropy functions\label{s:anisofun}}
These are given by
\begin{align}
   f(x,y) &=  -\int \frac{d\Omega_k}{4\pi}  \biggl(\frac{3\cos^2\theta_k}{\phi_+\sin^2\theta_k+\cos^2\theta_k}-1\biggr),\\
f_2(x,y) &=  \int \frac{d\Omega_k}{4\pi}\frac{\phi_-\sin^2\theta_k}{\phi_+\sin^2\theta_k+\cos^2\theta_k},
\end{align}
where $\phi_\pm\equiv x^{-2}\cos^2\phi_k\pm y^{-2}\sin^2\phi_k$. 

\section{Large $N$ limit\label{s:largeN}}
The droplet profile becomes increasingly flat topped so that $\int d\bx|\psi|^p\to \npeak^{p/2-1}N$, kinetic energy becomes insignificant, and we find $\npeak$ that minimizes 
\begin{align}
    \frac{E}{N} &\to  \frac12g_s(1-\elim)\npeak + \frac25\gammaQF \npeak^{3/2}. \label{e:EnoK}
\end{align}
%
%\bibliography{\string~/Dropbox/tex/bib/physjabbrev.bib,\string~/Dropbox/tex/bib/articles.bib,\string~/Dropbox/tex/bib/books.bib}

\begin{thebibliography}{63}%
\makeatletter
\providecommand \@ifxundefined [1]{%
 \@ifx{#1\undefined}
}%
\providecommand \@ifnum [1]{%
 \ifnum #1\expandafter \@firstoftwo
 \else \expandafter \@secondoftwo
 \fi
}%
\providecommand \@ifx [1]{%
 \ifx #1\expandafter \@firstoftwo
 \else \expandafter \@secondoftwo
 \fi
}%
\providecommand \natexlab [1]{#1}%
\providecommand \enquote  [1]{``#1''}%
\providecommand \bibnamefont  [1]{#1}%
\providecommand \bibfnamefont [1]{#1}%
\providecommand \citenamefont [1]{#1}%
\providecommand \href@noop [0]{\@secondoftwo}%
\providecommand \href [0]{\begingroup \@sanitize@url \@href}%
\providecommand \@href[1]{\@@startlink{#1}\@@href}%
\providecommand \@@href[1]{\endgroup#1\@@endlink}%
\providecommand \@sanitize@url [0]{\catcode `\\12\catcode `\$12\catcode
  `\&12\catcode `\#12\catcode `\^12\catcode `\_12\catcode `\%12\relax}%
\providecommand \@@startlink[1]{}%
\providecommand \@@endlink[0]{}%
\providecommand \url  [0]{\begingroup\@sanitize@url \@url }%
\providecommand \@url [1]{\endgroup\@href {#1}{\urlprefix }}%
\providecommand \urlprefix  [0]{URL }%
\providecommand \Eprint [0]{\href }%
\providecommand \doibase [0]{https://doi.org/}%
\providecommand \selectlanguage [0]{\@gobble}%
\providecommand \bibinfo  [0]{\@secondoftwo}%
\providecommand \bibfield  [0]{\@secondoftwo}%
\providecommand \translation [1]{[#1]}%
\providecommand \BibitemOpen [0]{}%
\providecommand \bibitemStop [0]{}%
\providecommand \bibitemNoStop [0]{.\EOS\space}%
\providecommand \EOS [0]{\spacefactor3000\relax}%
\providecommand \BibitemShut  [1]{\csname bibitem#1\endcsname}%
\let\auto@bib@innerbib\@empty
%</preamble>
\bibitem [{\citenamefont {Ni}\ \emph {et~al.}(2008)\citenamefont {Ni},
  \citenamefont {Ospelkaus}, \citenamefont {de~Miranda}, \citenamefont {Pe'er},
  \citenamefont {Neyenhuis}, \citenamefont {Zirbel}, \citenamefont
  {Kotochigova}, \citenamefont {Julienne}, \citenamefont {Jin},\ and\
  \citenamefont {Ye}}]{Ni2008a}%
  \BibitemOpen
  \bibfield  {author} {\bibinfo {author} {\bibfnamefont {K.~K.}\ \bibnamefont
  {Ni}}, \bibinfo {author} {\bibfnamefont {S.}~\bibnamefont {Ospelkaus}},
  \bibinfo {author} {\bibfnamefont {M.~H.~G.}\ \bibnamefont {de~Miranda}},
  \bibinfo {author} {\bibfnamefont {A.}~\bibnamefont {Pe'er}}, \bibinfo
  {author} {\bibfnamefont {B.}~\bibnamefont {Neyenhuis}}, \bibinfo {author}
  {\bibfnamefont {J.~J.}\ \bibnamefont {Zirbel}}, \bibinfo {author}
  {\bibfnamefont {S.}~\bibnamefont {Kotochigova}}, \bibinfo {author}
  {\bibfnamefont {P.~S.}\ \bibnamefont {Julienne}}, \bibinfo {author}
  {\bibfnamefont {D.~S.}\ \bibnamefont {Jin}},\ and\ \bibinfo {author}
  {\bibfnamefont {J.}~\bibnamefont {Ye}},\ }\bibfield  {title} {\bibinfo
  {title} {A high phase-space-density gas of polar molecules},\ }\href
  {https://doi.org/10.1126/science.1163861} {\bibfield  {journal} {\bibinfo
  {journal} {Science}\ }\textbf {\bibinfo {volume} {322}},\ \bibinfo {pages}
  {231} (\bibinfo {year} {2008})}\BibitemShut {NoStop}%
\bibitem [{\citenamefont {Ospelkaus}\ \emph {et~al.}(2010)\citenamefont
  {Ospelkaus}, \citenamefont {Ni}, \citenamefont {Wang}, \citenamefont
  {de~Miranda}, \citenamefont {Neyenhuis}, \citenamefont {Qu{\'e}m{\'e}ner},
  \citenamefont {Julienne}, \citenamefont {Bohn}, \citenamefont {Jin},\ and\
  \citenamefont {Ye}}]{Ospelkaus2010a}%
  \BibitemOpen
  \bibfield  {author} {\bibinfo {author} {\bibfnamefont {S.}~\bibnamefont
  {Ospelkaus}}, \bibinfo {author} {\bibfnamefont {K.-K.}\ \bibnamefont {Ni}},
  \bibinfo {author} {\bibfnamefont {D.}~\bibnamefont {Wang}}, \bibinfo {author}
  {\bibfnamefont {M.~H.~G.}\ \bibnamefont {de~Miranda}}, \bibinfo {author}
  {\bibfnamefont {B.}~\bibnamefont {Neyenhuis}}, \bibinfo {author}
  {\bibfnamefont {G.}~\bibnamefont {Qu{\'e}m{\'e}ner}}, \bibinfo {author}
  {\bibfnamefont {P.~S.}\ \bibnamefont {Julienne}}, \bibinfo {author}
  {\bibfnamefont {J.~L.}\ \bibnamefont {Bohn}}, \bibinfo {author}
  {\bibfnamefont {D.~S.}\ \bibnamefont {Jin}},\ and\ \bibinfo {author}
  {\bibfnamefont {J.}~\bibnamefont {Ye}},\ }\bibfield  {title} {\bibinfo
  {title} {Quantum-state controlled chemical reactions of ultracold
  potassium-rubidium molecules},\ }\href
  {https://doi.org/10.1126/science.1184121} {\bibfield  {journal} {\bibinfo
  {journal} {Science}\ }\textbf {\bibinfo {volume} {327}},\ \bibinfo {pages}
  {853} (\bibinfo {year} {2010})}\BibitemShut {NoStop}%
\bibitem [{\citenamefont {Ni}\ \emph {et~al.}(2010)\citenamefont {Ni},
  \citenamefont {Ospelkaus}, \citenamefont {Wang}, \citenamefont {Quemener},
  \citenamefont {Neyenhuis}, \citenamefont {de~Miranda}, \citenamefont {Bohn},
  \citenamefont {Ye},\ and\ \citenamefont {Jin}}]{Ni2010a}%
  \BibitemOpen
  \bibfield  {author} {\bibinfo {author} {\bibfnamefont {K.-K.}\ \bibnamefont
  {Ni}}, \bibinfo {author} {\bibfnamefont {S.}~\bibnamefont {Ospelkaus}},
  \bibinfo {author} {\bibfnamefont {D.}~\bibnamefont {Wang}}, \bibinfo {author}
  {\bibfnamefont {G.}~\bibnamefont {Quemener}}, \bibinfo {author}
  {\bibfnamefont {B.}~\bibnamefont {Neyenhuis}}, \bibinfo {author}
  {\bibfnamefont {M.~H.~G.}\ \bibnamefont {de~Miranda}}, \bibinfo {author}
  {\bibfnamefont {J.~L.}\ \bibnamefont {Bohn}}, \bibinfo {author}
  {\bibfnamefont {J.}~\bibnamefont {Ye}},\ and\ \bibinfo {author}
  {\bibfnamefont {D.~S.}\ \bibnamefont {Jin}},\ }\bibfield  {title} {\bibinfo
  {title} {Dipolar collisions of polar molecules in the quantum regime},\
  }\href {https://doi.org/10.1038/nature08953} {\bibfield  {journal} {\bibinfo
  {journal} {Nature}\ }\textbf {\bibinfo {volume} {464}},\ \bibinfo {pages}
  {1324} (\bibinfo {year} {2010})}\BibitemShut {NoStop}%
\bibitem [{\citenamefont {De~Marco}\ \emph {et~al.}(2019)\citenamefont
  {De~Marco}, \citenamefont {Valtolina}, \citenamefont {Matsuda}, \citenamefont
  {Tobias}, \citenamefont {Covey},\ and\ \citenamefont {Ye}}]{De-Marco2019a}%
  \BibitemOpen
  \bibfield  {author} {\bibinfo {author} {\bibfnamefont {L.}~\bibnamefont
  {De~Marco}}, \bibinfo {author} {\bibfnamefont {G.}~\bibnamefont {Valtolina}},
  \bibinfo {author} {\bibfnamefont {K.}~\bibnamefont {Matsuda}}, \bibinfo
  {author} {\bibfnamefont {W.~G.}\ \bibnamefont {Tobias}}, \bibinfo {author}
  {\bibfnamefont {J.~P.}\ \bibnamefont {Covey}},\ and\ \bibinfo {author}
  {\bibfnamefont {J.}~\bibnamefont {Ye}},\ }\bibfield  {title} {\bibinfo
  {title} {A degenerate {Fermi} gas of polar molecules},\ }\href
  {https://doi.org/10.1126/science.aau7230} {\bibfield  {journal} {\bibinfo
  {journal} {Science}\ }\textbf {\bibinfo {volume} {363}},\ \bibinfo {pages}
  {853} (\bibinfo {year} {2019})}\BibitemShut {NoStop}%
\bibitem [{\citenamefont {Karman}\ and\ \citenamefont
  {Hutson}(2018)}]{Karman2018a}%
  \BibitemOpen
  \bibfield  {author} {\bibinfo {author} {\bibfnamefont {T.}~\bibnamefont
  {Karman}}\ and\ \bibinfo {author} {\bibfnamefont {J.~M.}\ \bibnamefont
  {Hutson}},\ }\bibfield  {title} {\bibinfo {title} {Microwave shielding of
  ultracold polar molecules},\ }\href
  {https://doi.org/10.1103/PhysRevLett.121.163401} {\bibfield  {journal}
  {\bibinfo  {journal} {Phys. Rev. Lett.}\ }\textbf {\bibinfo {volume} {121}},\
  \bibinfo {pages} {163401} (\bibinfo {year} {2018})}\BibitemShut {NoStop}%
\bibitem [{\citenamefont {Lassabli\`ere}\ and\ \citenamefont
  {Qu\'em\'ener}(2018)}]{Lassabliere2018a}%
  \BibitemOpen
  \bibfield  {author} {\bibinfo {author} {\bibfnamefont {L.}~\bibnamefont
  {Lassabli\`ere}}\ and\ \bibinfo {author} {\bibfnamefont {G.}~\bibnamefont
  {Qu\'em\'ener}},\ }\bibfield  {title} {\bibinfo {title} {Controlling the
  scattering length of ultracold dipolar molecules},\ }\href
  {https://doi.org/10.1103/PhysRevLett.121.163402} {\bibfield  {journal}
  {\bibinfo  {journal} {Phys. Rev. Lett.}\ }\textbf {\bibinfo {volume} {121}},\
  \bibinfo {pages} {163402} (\bibinfo {year} {2018})}\BibitemShut {NoStop}%
\bibitem [{\citenamefont {Anderegg}\ \emph {et~al.}(2021)\citenamefont
  {Anderegg}, \citenamefont {Burchesky}, \citenamefont {Bao}, \citenamefont
  {Yu}, \citenamefont {Karman}, \citenamefont {Chae}, \citenamefont {Ni},
  \citenamefont {Ketterle},\ and\ \citenamefont {Doyle}}]{Anderegg2021a}%
  \BibitemOpen
  \bibfield  {author} {\bibinfo {author} {\bibfnamefont {L.}~\bibnamefont
  {Anderegg}}, \bibinfo {author} {\bibfnamefont {S.}~\bibnamefont {Burchesky}},
  \bibinfo {author} {\bibfnamefont {Y.}~\bibnamefont {Bao}}, \bibinfo {author}
  {\bibfnamefont {S.~S.}\ \bibnamefont {Yu}}, \bibinfo {author} {\bibfnamefont
  {T.}~\bibnamefont {Karman}}, \bibinfo {author} {\bibfnamefont
  {E.}~\bibnamefont {Chae}}, \bibinfo {author} {\bibfnamefont {K.-K.}\
  \bibnamefont {Ni}}, \bibinfo {author} {\bibfnamefont {W.}~\bibnamefont
  {Ketterle}},\ and\ \bibinfo {author} {\bibfnamefont {J.~M.}\ \bibnamefont
  {Doyle}},\ }\bibfield  {title} {\bibinfo {title} {Observation of microwave
  shielding of ultracold molecules},\ }\href
  {https://doi.org/10.1126/science.abg9502} {\bibfield  {journal} {\bibinfo
  {journal} {Science}\ }\textbf {\bibinfo {volume} {373}},\ \bibinfo {pages}
  {779} (\bibinfo {year} {2021})}\BibitemShut {NoStop}%
\bibitem [{\citenamefont {Bigagli}\ \emph {et~al.}(2023)\citenamefont
  {Bigagli}, \citenamefont {Warner}, \citenamefont {Yuan}, \citenamefont
  {Zhang}, \citenamefont {Stevenson}, \citenamefont {Karman},\ and\
  \citenamefont {Will}}]{Bigagli2023a}%
  \BibitemOpen
  \bibfield  {author} {\bibinfo {author} {\bibfnamefont {N.}~\bibnamefont
  {Bigagli}}, \bibinfo {author} {\bibfnamefont {C.}~\bibnamefont {Warner}},
  \bibinfo {author} {\bibfnamefont {W.}~\bibnamefont {Yuan}}, \bibinfo {author}
  {\bibfnamefont {S.}~\bibnamefont {Zhang}}, \bibinfo {author} {\bibfnamefont
  {I.}~\bibnamefont {Stevenson}}, \bibinfo {author} {\bibfnamefont
  {T.}~\bibnamefont {Karman}},\ and\ \bibinfo {author} {\bibfnamefont
  {S.}~\bibnamefont {Will}},\ }\bibfield  {title} {\bibinfo {title}
  {Collisionally stable gas of bosonic dipolar ground-state molecules},\ }\href
  {https://doi.org/10.1038/s41567-023-02200-6} {\bibfield  {journal} {\bibinfo
  {journal} {Nature Physics}\ }\textbf {\bibinfo {volume} {19}},\ \bibinfo
  {pages} {1579} (\bibinfo {year} {2023})}\BibitemShut {NoStop}%
\bibitem [{\citenamefont {Bigagli}\ \emph {et~al.}(2024)\citenamefont
  {Bigagli}, \citenamefont {Yuan}, \citenamefont {Zhang}, \citenamefont
  {Bulatovic}, \citenamefont {Karman}, \citenamefont {Stevenson},\ and\
  \citenamefont {Will}}]{Bigagli2024a}%
  \BibitemOpen
  \bibfield  {author} {\bibinfo {author} {\bibfnamefont {N.}~\bibnamefont
  {Bigagli}}, \bibinfo {author} {\bibfnamefont {W.}~\bibnamefont {Yuan}},
  \bibinfo {author} {\bibfnamefont {S.}~\bibnamefont {Zhang}}, \bibinfo
  {author} {\bibfnamefont {B.}~\bibnamefont {Bulatovic}}, \bibinfo {author}
  {\bibfnamefont {T.}~\bibnamefont {Karman}}, \bibinfo {author} {\bibfnamefont
  {I.}~\bibnamefont {Stevenson}},\ and\ \bibinfo {author} {\bibfnamefont
  {S.}~\bibnamefont {Will}},\ }\bibfield  {title} {\bibinfo {title}
  {Observation of {Bose}--{Einstein} condensation of dipolar molecules},\
  }\href {https://doi.org/10.1038/s41586-024-07492-z} {\bibfield  {journal}
  {\bibinfo  {journal} {Nature}\ }\textbf {\bibinfo {volume} {631}},\ \bibinfo
  {pages} {289} (\bibinfo {year} {2024})}\BibitemShut {NoStop}%
\bibitem [{\citenamefont {Langen}\ \emph {et~al.}(2024)\citenamefont {Langen},
  \citenamefont {Valtolina}, \citenamefont {Wang},\ and\ \citenamefont
  {Ye}}]{Langen2024a}%
  \BibitemOpen
  \bibfield  {author} {\bibinfo {author} {\bibfnamefont {T.}~\bibnamefont
  {Langen}}, \bibinfo {author} {\bibfnamefont {G.}~\bibnamefont {Valtolina}},
  \bibinfo {author} {\bibfnamefont {D.}~\bibnamefont {Wang}},\ and\ \bibinfo
  {author} {\bibfnamefont {J.}~\bibnamefont {Ye}},\ }\bibfield  {title}
  {\bibinfo {title} {Quantum state manipulation and cooling of ultracold
  molecules},\ }\href {https://doi.org/10.1038/s41567-024-02423-1} {\bibfield
  {journal} {\bibinfo  {journal} {Nature Physics}\ }\textbf {\bibinfo {volume}
  {20}},\ \bibinfo {pages} {702} (\bibinfo {year} {2024})}\BibitemShut
  {NoStop}%
\bibitem [{\citenamefont {Lin}\ \emph {et~al.}(2023)\citenamefont {Lin},
  \citenamefont {Chen}, \citenamefont {Jin}, \citenamefont {Shi}, \citenamefont
  {Deng}, \citenamefont {Zhang}, \citenamefont {Qu\'em\'ener}, \citenamefont
  {Shi}, \citenamefont {Yi},\ and\ \citenamefont {Wang}}]{Lin2023a}%
  \BibitemOpen
  \bibfield  {author} {\bibinfo {author} {\bibfnamefont {J.}~\bibnamefont
  {Lin}}, \bibinfo {author} {\bibfnamefont {G.}~\bibnamefont {Chen}}, \bibinfo
  {author} {\bibfnamefont {M.}~\bibnamefont {Jin}}, \bibinfo {author}
  {\bibfnamefont {Z.}~\bibnamefont {Shi}}, \bibinfo {author} {\bibfnamefont
  {F.}~\bibnamefont {Deng}}, \bibinfo {author} {\bibfnamefont {W.}~\bibnamefont
  {Zhang}}, \bibinfo {author} {\bibfnamefont {G.}~\bibnamefont {Qu\'em\'ener}},
  \bibinfo {author} {\bibfnamefont {T.}~\bibnamefont {Shi}}, \bibinfo {author}
  {\bibfnamefont {S.}~\bibnamefont {Yi}},\ and\ \bibinfo {author}
  {\bibfnamefont {D.}~\bibnamefont {Wang}},\ }\bibfield  {title} {\bibinfo
  {title} {Microwave shielding of bosonic {NaRb} molecules},\ }\href
  {https://doi.org/10.1103/PhysRevX.13.031032} {\bibfield  {journal} {\bibinfo
  {journal} {Phys. Rev. X}\ }\textbf {\bibinfo {volume} {13}},\ \bibinfo
  {pages} {031032} (\bibinfo {year} {2023})}\BibitemShut {NoStop}%
\bibitem [{\citenamefont {Shi}\ \emph {et~al.}(2025)\citenamefont {Shi},
  \citenamefont {Huang}, \citenamefont {Deng}, \citenamefont {Jin},
  \citenamefont {Yi}, \citenamefont {Shi},\ and\ \citenamefont
  {Wang}}]{Shi2025a}%
  \BibitemOpen
  \bibfield  {author} {\bibinfo {author} {\bibfnamefont {Z.}~\bibnamefont
  {Shi}}, \bibinfo {author} {\bibfnamefont {Z.}~\bibnamefont {Huang}}, \bibinfo
  {author} {\bibfnamefont {F.}~\bibnamefont {Deng}}, \bibinfo {author}
  {\bibfnamefont {W.-J.}\ \bibnamefont {Jin}}, \bibinfo {author} {\bibfnamefont
  {S.}~\bibnamefont {Yi}}, \bibinfo {author} {\bibfnamefont {T.}~\bibnamefont
  {Shi}},\ and\ \bibinfo {author} {\bibfnamefont {D.}~\bibnamefont {Wang}},\
  }\href {https://arxiv.org/abs/2508.20518} {\bibinfo {title} {Bose-{Einstein}
  condensate of ultracold sodium-rubidium molecules with tunable dipolar
  interactions}} (\bibinfo {year} {2025}),\ \Eprint
  {https://arxiv.org/abs/2508.20518} {arXiv:2508.20518} \BibitemShut {NoStop}%
\bibitem [{\citenamefont {Kadau}\ \emph {et~al.}(2016)\citenamefont {Kadau},
  \citenamefont {Schmitt}, \citenamefont {Wenzel}, \citenamefont {Wink},
  \citenamefont {Maier}, \citenamefont {Ferrier-Barbut},\ and\ \citenamefont
  {Pfau}}]{Kadau2016a}%
  \BibitemOpen
  \bibfield  {author} {\bibinfo {author} {\bibfnamefont {H.}~\bibnamefont
  {Kadau}}, \bibinfo {author} {\bibfnamefont {M.}~\bibnamefont {Schmitt}},
  \bibinfo {author} {\bibfnamefont {M.}~\bibnamefont {Wenzel}}, \bibinfo
  {author} {\bibfnamefont {C.}~\bibnamefont {Wink}}, \bibinfo {author}
  {\bibfnamefont {T.}~\bibnamefont {Maier}}, \bibinfo {author} {\bibfnamefont
  {I.}~\bibnamefont {Ferrier-Barbut}},\ and\ \bibinfo {author} {\bibfnamefont
  {T.}~\bibnamefont {Pfau}},\ }\bibfield  {title} {\bibinfo {title} {Observing
  the {Rosensweig} instability of a quantum ferrofluid},\ }\href
  {http://dx.doi.org/10.1038/nature16485} {\bibfield  {journal} {\bibinfo
  {journal} {Nature}\ }\textbf {\bibinfo {volume} {530}},\ \bibinfo {pages}
  {194} (\bibinfo {year} {2016})}\BibitemShut {NoStop}%
\bibitem [{\citenamefont {Chomaz}\ \emph {et~al.}(2016)\citenamefont {Chomaz},
  \citenamefont {Baier}, \citenamefont {Petter}, \citenamefont {Mark},
  \citenamefont {W\"achtler}, \citenamefont {Santos},\ and\ \citenamefont
  {Ferlaino}}]{Chomaz2016a}%
  \BibitemOpen
  \bibfield  {author} {\bibinfo {author} {\bibfnamefont {L.}~\bibnamefont
  {Chomaz}}, \bibinfo {author} {\bibfnamefont {S.}~\bibnamefont {Baier}},
  \bibinfo {author} {\bibfnamefont {D.}~\bibnamefont {Petter}}, \bibinfo
  {author} {\bibfnamefont {M.~J.}\ \bibnamefont {Mark}}, \bibinfo {author}
  {\bibfnamefont {F.}~\bibnamefont {W\"achtler}}, \bibinfo {author}
  {\bibfnamefont {L.}~\bibnamefont {Santos}},\ and\ \bibinfo {author}
  {\bibfnamefont {F.}~\bibnamefont {Ferlaino}},\ }\bibfield  {title} {\bibinfo
  {title} {Quantum-fluctuation-driven crossover from a dilute {Bose}-{Einstein}
  condensate to a macrodroplet in a dipolar quantum fluid},\ }\href
  {https://doi.org/10.1103/PhysRevX.6.041039} {\bibfield  {journal} {\bibinfo
  {journal} {Phys. Rev. X}\ }\textbf {\bibinfo {volume} {6}},\ \bibinfo {pages}
  {041039} (\bibinfo {year} {2016})}\BibitemShut {NoStop}%
\bibitem [{\citenamefont {Schmitt}\ \emph {et~al.}(2016)\citenamefont
  {Schmitt}, \citenamefont {Wenzel}, \citenamefont {B{\"o}ttcher},
  \citenamefont {Ferrier-Barbut},\ and\ \citenamefont {Pfau}}]{Schmitt2016a}%
  \BibitemOpen
  \bibfield  {author} {\bibinfo {author} {\bibfnamefont {M.}~\bibnamefont
  {Schmitt}}, \bibinfo {author} {\bibfnamefont {M.}~\bibnamefont {Wenzel}},
  \bibinfo {author} {\bibfnamefont {F.}~\bibnamefont {B{\"o}ttcher}}, \bibinfo
  {author} {\bibfnamefont {I.}~\bibnamefont {Ferrier-Barbut}},\ and\ \bibinfo
  {author} {\bibfnamefont {T.}~\bibnamefont {Pfau}},\ }\bibfield  {title}
  {\bibinfo {title} {Self-bound droplets of a dilute magnetic quantum liquid},\
  }\href {http://dx.doi.org/10.1038/nature20126} {\bibfield  {journal}
  {\bibinfo  {journal} {Nature}\ }\textbf {\bibinfo {volume} {539}},\ \bibinfo
  {pages} {259} (\bibinfo {year} {2016})}\BibitemShut {NoStop}%
\bibitem [{\citenamefont {Petrov}(2015)}]{Petrov2015a}%
  \BibitemOpen
  \bibfield  {author} {\bibinfo {author} {\bibfnamefont {D.~S.}\ \bibnamefont
  {Petrov}},\ }\bibfield  {title} {\bibinfo {title} {Quantum mechanical
  stabilization of a collapsing {Bose}-{Bose} mixture},\ }\href
  {https://doi.org/10.1103/PhysRevLett.115.155302} {\bibfield  {journal}
  {\bibinfo  {journal} {Phys. Rev. Lett.}\ }\textbf {\bibinfo {volume} {115}},\
  \bibinfo {pages} {155302} (\bibinfo {year} {2015})}\BibitemShut {NoStop}%
\bibitem [{\citenamefont {Lahaye}\ \emph {et~al.}(2009)\citenamefont {Lahaye},
  \citenamefont {Menotti}, \citenamefont {Santos}, \citenamefont {Lewenstein},\
  and\ \citenamefont {Pfau}}]{Lahaye2009a}%
  \BibitemOpen
  \bibfield  {author} {\bibinfo {author} {\bibfnamefont {T.}~\bibnamefont
  {Lahaye}}, \bibinfo {author} {\bibfnamefont {C.}~\bibnamefont {Menotti}},
  \bibinfo {author} {\bibfnamefont {L.}~\bibnamefont {Santos}}, \bibinfo
  {author} {\bibfnamefont {M.}~\bibnamefont {Lewenstein}},\ and\ \bibinfo
  {author} {\bibfnamefont {T.}~\bibnamefont {Pfau}},\ }\bibfield  {title}
  {\bibinfo {title} {The physics of dipolar bosonic quantum gases},\ }\href
  {http://stacks.iop.org/0034-4885/72/126401} {\bibfield  {journal} {\bibinfo
  {journal} {Rep. Prog. Phys.}\ }\textbf {\bibinfo {volume} {72}},\ \bibinfo
  {pages} {126401} (\bibinfo {year} {2009})}\BibitemShut {NoStop}%
\bibitem [{\citenamefont {Chomaz}\ \emph {et~al.}(2022)\citenamefont {Chomaz},
  \citenamefont {Ferrier-Barbut}, \citenamefont {Ferlaino}, \citenamefont
  {Laburthe-Tolra}, \citenamefont {Lev},\ and\ \citenamefont
  {Pfau}}]{Chomaz2023a}%
  \BibitemOpen
  \bibfield  {author} {\bibinfo {author} {\bibfnamefont {L.}~\bibnamefont
  {Chomaz}}, \bibinfo {author} {\bibfnamefont {I.}~\bibnamefont
  {Ferrier-Barbut}}, \bibinfo {author} {\bibfnamefont {F.}~\bibnamefont
  {Ferlaino}}, \bibinfo {author} {\bibfnamefont {B.}~\bibnamefont
  {Laburthe-Tolra}}, \bibinfo {author} {\bibfnamefont {B.~L.}\ \bibnamefont
  {Lev}},\ and\ \bibinfo {author} {\bibfnamefont {T.}~\bibnamefont {Pfau}},\
  }\bibfield  {title} {\bibinfo {title} {Dipolar physics: a review of
  experiments with magnetic quantum gases},\ }\href
  {https://doi.org/10.1088/1361-6633/aca814} {\bibfield  {journal} {\bibinfo
  {journal} {Rep. Prog. Phys.}\ }\textbf {\bibinfo {volume} {86}},\ \bibinfo
  {pages} {026401} (\bibinfo {year} {2022})}\BibitemShut {NoStop}%
\bibitem [{\citenamefont {Chomaz}(2026)}]{Chomaz2026a}%
  \BibitemOpen
  \bibfield  {author} {\bibinfo {author} {\bibfnamefont {L.}~\bibnamefont
  {Chomaz}},\ }\bibfield  {title} {\bibinfo {title} {Quantum-stabilized states
  in magnetic dipolar quantum gases},\ }\href
  {https://doi.org/https://doi.org/10.1146/annurev-conmatphys-061125-032048}
  {\bibfield  {journal} {\bibinfo  {journal} {Annu. Rev. Condens. Matter
  Phys.}\ }\textbf {\bibinfo {volume} {17}},\ \bibinfo {pages} {47} (\bibinfo
  {year} {2026})}\BibitemShut {NoStop}%
\bibitem [{\citenamefont {Baillie}\ \emph {et~al.}(2016)\citenamefont
  {Baillie}, \citenamefont {Wilson}, \citenamefont {Bisset},\ and\
  \citenamefont {Blakie}}]{Baillie2016b}%
  \BibitemOpen
  \bibfield  {author} {\bibinfo {author} {\bibfnamefont {D.}~\bibnamefont
  {Baillie}}, \bibinfo {author} {\bibfnamefont {R.~M.}\ \bibnamefont {Wilson}},
  \bibinfo {author} {\bibfnamefont {R.~N.}\ \bibnamefont {Bisset}},\ and\
  \bibinfo {author} {\bibfnamefont {P.~B.}\ \bibnamefont {Blakie}},\ }\bibfield
   {title} {\bibinfo {title} {Self-bound dipolar droplet: A localized matter
  wave in free space},\ }\href {https://doi.org/10.1103/PhysRevA.94.021602}
  {\bibfield  {journal} {\bibinfo  {journal} {Phys. Rev. A}\ }\textbf {\bibinfo
  {volume} {94}},\ \bibinfo {pages} {021602(R)} (\bibinfo {year}
  {2016})}\BibitemShut {NoStop}%
\bibitem [{\citenamefont {Bisset}\ \emph {et~al.}(2016)\citenamefont {Bisset},
  \citenamefont {Wilson}, \citenamefont {Baillie},\ and\ \citenamefont
  {Blakie}}]{Bisset2016a}%
  \BibitemOpen
  \bibfield  {author} {\bibinfo {author} {\bibfnamefont {R.~N.}\ \bibnamefont
  {Bisset}}, \bibinfo {author} {\bibfnamefont {R.~M.}\ \bibnamefont {Wilson}},
  \bibinfo {author} {\bibfnamefont {D.}~\bibnamefont {Baillie}},\ and\ \bibinfo
  {author} {\bibfnamefont {P.~B.}\ \bibnamefont {Blakie}},\ }\bibfield  {title}
  {\bibinfo {title} {Ground-state phase diagram of a dipolar condensate with
  quantum fluctuations},\ }\href {https://doi.org/10.1103/PhysRevA.94.033619}
  {\bibfield  {journal} {\bibinfo  {journal} {Phys. Rev. A}\ }\textbf {\bibinfo
  {volume} {94}},\ \bibinfo {pages} {033619} (\bibinfo {year}
  {2016})}\BibitemShut {NoStop}%
\bibitem [{\citenamefont {W\"achtler}\ and\ \citenamefont
  {Santos}(2016{\natexlab{a}})}]{Wachtler2016a}%
  \BibitemOpen
  \bibfield  {author} {\bibinfo {author} {\bibfnamefont {F.}~\bibnamefont
  {W\"achtler}}\ and\ \bibinfo {author} {\bibfnamefont {L.}~\bibnamefont
  {Santos}},\ }\bibfield  {title} {\bibinfo {title} {Quantum filaments in
  dipolar {Bose}-{Einstein} condensates},\ }\href
  {https://doi.org/10.1103/PhysRevA.93.061603} {\bibfield  {journal} {\bibinfo
  {journal} {Phys. Rev. A}\ }\textbf {\bibinfo {volume} {93}},\ \bibinfo
  {pages} {061603} (\bibinfo {year} {2016}{\natexlab{a}})}\BibitemShut
  {NoStop}%
\bibitem [{\citenamefont {W\"achtler}\ and\ \citenamefont
  {Santos}(2016{\natexlab{b}})}]{Wachtler2016b}%
  \BibitemOpen
  \bibfield  {author} {\bibinfo {author} {\bibfnamefont {F.}~\bibnamefont
  {W\"achtler}}\ and\ \bibinfo {author} {\bibfnamefont {L.}~\bibnamefont
  {Santos}},\ }\bibfield  {title} {\bibinfo {title} {Ground-state properties
  and elementary excitations of quantum droplets in dipolar {Bose}-{Einstein}
  condensates},\ }\href {https://doi.org/10.1103/PhysRevA.94.043618} {\bibfield
   {journal} {\bibinfo  {journal} {Phys. Rev. A}\ }\textbf {\bibinfo {volume}
  {94}},\ \bibinfo {pages} {043618} (\bibinfo {year}
  {2016}{\natexlab{b}})}\BibitemShut {NoStop}%
\bibitem [{\citenamefont {Cabrera}\ \emph {et~al.}(2018)\citenamefont
  {Cabrera}, \citenamefont {Tanzi}, \citenamefont {Sanz}, \citenamefont
  {Naylor}, \citenamefont {Thomas}, \citenamefont {Cheiney},\ and\
  \citenamefont {Tarruell}}]{Cabrera2018a}%
  \BibitemOpen
  \bibfield  {author} {\bibinfo {author} {\bibfnamefont {C.~R.}\ \bibnamefont
  {Cabrera}}, \bibinfo {author} {\bibfnamefont {L.}~\bibnamefont {Tanzi}},
  \bibinfo {author} {\bibfnamefont {J.}~\bibnamefont {Sanz}}, \bibinfo {author}
  {\bibfnamefont {B.}~\bibnamefont {Naylor}}, \bibinfo {author} {\bibfnamefont
  {P.}~\bibnamefont {Thomas}}, \bibinfo {author} {\bibfnamefont
  {P.}~\bibnamefont {Cheiney}},\ and\ \bibinfo {author} {\bibfnamefont
  {L.}~\bibnamefont {Tarruell}},\ }\bibfield  {title} {\bibinfo {title}
  {Quantum liquid droplets in a mixture of {Bose}-{Einstein} condensates},\
  }\href {https://doi.org/10.1126/science.aao5686} {\bibfield  {journal}
  {\bibinfo  {journal} {Science}\ }\textbf {\bibinfo {volume} {359}},\ \bibinfo
  {pages} {301} (\bibinfo {year} {2018})}\BibitemShut {NoStop}%
\bibitem [{\citenamefont {Semeghini}\ \emph {et~al.}(2018)\citenamefont
  {Semeghini}, \citenamefont {Ferioli}, \citenamefont {Masi}, \citenamefont
  {Mazzinghi}, \citenamefont {Wolswijk}, \citenamefont {Minardi}, \citenamefont
  {Modugno}, \citenamefont {Modugno}, \citenamefont {Inguscio},\ and\
  \citenamefont {Fattori}}]{Semeghini2018a}%
  \BibitemOpen
  \bibfield  {author} {\bibinfo {author} {\bibfnamefont {G.}~\bibnamefont
  {Semeghini}}, \bibinfo {author} {\bibfnamefont {G.}~\bibnamefont {Ferioli}},
  \bibinfo {author} {\bibfnamefont {L.}~\bibnamefont {Masi}}, \bibinfo {author}
  {\bibfnamefont {C.}~\bibnamefont {Mazzinghi}}, \bibinfo {author}
  {\bibfnamefont {L.}~\bibnamefont {Wolswijk}}, \bibinfo {author}
  {\bibfnamefont {F.}~\bibnamefont {Minardi}}, \bibinfo {author} {\bibfnamefont
  {M.}~\bibnamefont {Modugno}}, \bibinfo {author} {\bibfnamefont
  {G.}~\bibnamefont {Modugno}}, \bibinfo {author} {\bibfnamefont
  {M.}~\bibnamefont {Inguscio}},\ and\ \bibinfo {author} {\bibfnamefont
  {M.}~\bibnamefont {Fattori}},\ }\bibfield  {title} {\bibinfo {title}
  {Self-bound quantum droplets of atomic mixtures in free space},\ }\href
  {https://doi.org/10.1103/PhysRevLett.120.235301} {\bibfield  {journal}
  {\bibinfo  {journal} {Phys. Rev. Lett.}\ }\textbf {\bibinfo {volume} {120}},\
  \bibinfo {pages} {235301} (\bibinfo {year} {2018})}\BibitemShut {NoStop}%
\bibitem [{\citenamefont {Zhang}\ \emph {et~al.}(2026)\citenamefont {Zhang},
  \citenamefont {Yuan}, \citenamefont {Bigagli}, \citenamefont {Kwak},
  \citenamefont {Karman}, \citenamefont {Stevenson},\ and\ \citenamefont
  {Will}}]{Zhang2026a}%
  \BibitemOpen
  \bibfield  {author} {\bibinfo {author} {\bibfnamefont {S.}~\bibnamefont
  {Zhang}}, \bibinfo {author} {\bibfnamefont {W.}~\bibnamefont {Yuan}},
  \bibinfo {author} {\bibfnamefont {N.}~\bibnamefont {Bigagli}}, \bibinfo
  {author} {\bibfnamefont {H.}~\bibnamefont {Kwak}}, \bibinfo {author}
  {\bibfnamefont {T.}~\bibnamefont {Karman}}, \bibinfo {author} {\bibfnamefont
  {I.}~\bibnamefont {Stevenson}},\ and\ \bibinfo {author} {\bibfnamefont
  {S.}~\bibnamefont {Will}},\ }\bibfield  {title} {\bibinfo {title}
  {Observation of self-bound droplets of ultracold dipolar molecules},\ }\href
  {https://doi.org/10.1038/s41586-026-10245-9} {\bibfield  {journal} {\bibinfo
  {journal} {Nature}\ }\textbf {\bibinfo {volume} {651}},\ \bibinfo {pages}
  {601} (\bibinfo {year} {2026})}\BibitemShut {NoStop}%
\bibitem [{\citenamefont {Giovanazzi}\ \emph {et~al.}(2002)\citenamefont
  {Giovanazzi}, \citenamefont {G\"orlitz},\ and\ \citenamefont
  {Pfau}}]{Giovanazzi2002a}%
  \BibitemOpen
  \bibfield  {author} {\bibinfo {author} {\bibfnamefont {S.}~\bibnamefont
  {Giovanazzi}}, \bibinfo {author} {\bibfnamefont {A.}~\bibnamefont
  {G\"orlitz}},\ and\ \bibinfo {author} {\bibfnamefont {T.}~\bibnamefont
  {Pfau}},\ }\bibfield  {title} {\bibinfo {title} {Tuning the dipolar
  interaction in quantum gases},\ }\href
  {https://doi.org/10.1103/PhysRevLett.89.130401} {\bibfield  {journal}
  {\bibinfo  {journal} {Phys. Rev. Lett.}\ }\textbf {\bibinfo {volume} {89}},\
  \bibinfo {pages} {130401} (\bibinfo {year} {2002})}\BibitemShut {NoStop}%
\bibitem [{\citenamefont {Baillie}\ and\ \citenamefont
  {Blakie}(2020)}]{Baillie2020a}%
  \BibitemOpen
  \bibfield  {author} {\bibinfo {author} {\bibfnamefont {D.}~\bibnamefont
  {Baillie}}\ and\ \bibinfo {author} {\bibfnamefont {P.~B.}\ \bibnamefont
  {Blakie}},\ }\bibfield  {title} {\bibinfo {title} {Rotational tuning of the
  dipole-dipole interaction in a {Bose} gas of magnetic atoms},\ }\href
  {https://doi.org/10.1103/PhysRevA.101.043606} {\bibfield  {journal} {\bibinfo
   {journal} {Phys. Rev. A}\ }\textbf {\bibinfo {volume} {101}},\ \bibinfo
  {pages} {043606} (\bibinfo {year} {2020})}\BibitemShut {NoStop}%
\bibitem [{\citenamefont {Tang}\ \emph {et~al.}(2018)\citenamefont {Tang},
  \citenamefont {Kao}, \citenamefont {Li},\ and\ \citenamefont
  {Lev}}]{Tang2018a}%
  \BibitemOpen
  \bibfield  {author} {\bibinfo {author} {\bibfnamefont {Y.}~\bibnamefont
  {Tang}}, \bibinfo {author} {\bibfnamefont {W.}~\bibnamefont {Kao}}, \bibinfo
  {author} {\bibfnamefont {K.-Y.}\ \bibnamefont {Li}},\ and\ \bibinfo {author}
  {\bibfnamefont {B.~L.}\ \bibnamefont {Lev}},\ }\bibfield  {title} {\bibinfo
  {title} {Tuning the dipole-dipole interaction in a quantum gas with a
  rotating magnetic field},\ }\href
  {https://doi.org/10.1103/PhysRevLett.120.230401} {\bibfield  {journal}
  {\bibinfo  {journal} {Phys. Rev. Lett.}\ }\textbf {\bibinfo {volume} {120}},\
  \bibinfo {pages} {230401} (\bibinfo {year} {2018})}\BibitemShut {NoStop}%
\bibitem [{\citenamefont {Wenzel}(2018)}]{Wenzel2018c}%
  \BibitemOpen
  \bibfield  {author} {\bibinfo {author} {\bibfnamefont {M.}~\bibnamefont
  {Wenzel}},\ }\emph {\bibinfo {title} {Macroscopic states of dipolar quantum
  gases}},\ \href@noop {} {Ph.D. thesis},\ \bibinfo  {school} {University of
  Stuttgart}, \bibinfo {address} {Germany} (\bibinfo {year} {2018})\BibitemShut
  {NoStop}%
\bibitem [{\citenamefont {Mukherjee}\ \emph
  {et~al.}(2023{\natexlab{a}})\citenamefont {Mukherjee}, \citenamefont
  {Cardinale}, \citenamefont {Chergui}, \citenamefont {St{\"u}rmer},\ and\
  \citenamefont {Reimann}}]{Mukherjee2023b}%
  \BibitemOpen
  \bibfield  {author} {\bibinfo {author} {\bibfnamefont {K.}~\bibnamefont
  {Mukherjee}}, \bibinfo {author} {\bibfnamefont {T.~A.}\ \bibnamefont
  {Cardinale}}, \bibinfo {author} {\bibfnamefont {L.}~\bibnamefont {Chergui}},
  \bibinfo {author} {\bibfnamefont {P.}~\bibnamefont {St{\"u}rmer}},\ and\
  \bibinfo {author} {\bibfnamefont {S.~M.}\ \bibnamefont {Reimann}},\
  }\bibfield  {title} {\bibinfo {title} {Droplets and supersolids in ultra-cold
  atomic quantum gases},\ }\href
  {https://doi.org/10.1140/epjs/s11734-023-00991-6} {\bibfield  {journal}
  {\bibinfo  {journal} {EPJ ST}\ }\textbf {\bibinfo {volume} {232}},\ \bibinfo
  {pages} {3417} (\bibinfo {year} {2023}{\natexlab{a}})}\BibitemShut {NoStop}%
\bibitem [{\citenamefont {Mukherjee}\ \emph
  {et~al.}(2023{\natexlab{b}})\citenamefont {Mukherjee}, \citenamefont
  {Tengstrand}, \citenamefont {Cardinale},\ and\ \citenamefont
  {Reimann}}]{Mukherjee2023c}%
  \BibitemOpen
  \bibfield  {author} {\bibinfo {author} {\bibfnamefont {K.}~\bibnamefont
  {Mukherjee}}, \bibinfo {author} {\bibfnamefont {M.~N.}\ \bibnamefont
  {Tengstrand}}, \bibinfo {author} {\bibfnamefont {T.~A.}\ \bibnamefont
  {Cardinale}},\ and\ \bibinfo {author} {\bibfnamefont {S.~M.}\ \bibnamefont
  {Reimann}},\ }\bibfield  {title} {\bibinfo {title} {Supersolid stacks in
  antidipolar {Bose}-{Einstein} condensates},\ }\href
  {https://doi.org/10.1103/PhysRevA.108.023302} {\bibfield  {journal} {\bibinfo
   {journal} {Phys. Rev. A}\ }\textbf {\bibinfo {volume} {108}},\ \bibinfo
  {pages} {023302} (\bibinfo {year} {2023}{\natexlab{b}})}\BibitemShut
  {NoStop}%
\bibitem [{\citenamefont {Jin}\ \emph {et~al.}(2025)\citenamefont {Jin},
  \citenamefont {Deng}, \citenamefont {Yi},\ and\ \citenamefont
  {Shi}}]{Jin2025a}%
  \BibitemOpen
  \bibfield  {author} {\bibinfo {author} {\bibfnamefont {W.-J.}\ \bibnamefont
  {Jin}}, \bibinfo {author} {\bibfnamefont {F.}~\bibnamefont {Deng}}, \bibinfo
  {author} {\bibfnamefont {S.}~\bibnamefont {Yi}},\ and\ \bibinfo {author}
  {\bibfnamefont {T.}~\bibnamefont {Shi}},\ }\bibfield  {title} {\bibinfo
  {title} {Bose-{Einstein} condensates of microwave-shielded polar molecules},\
  }\href {https://doi.org/10.1103/b8y9-yvz9} {\bibfield  {journal} {\bibinfo
  {journal} {Phys. Rev. Lett.}\ }\textbf {\bibinfo {volume} {134}},\ \bibinfo
  {pages} {233003} (\bibinfo {year} {2025})}\BibitemShut {NoStop}%
\bibitem [{\citenamefont {Zhang}\ \emph
  {et~al.}(2025{\natexlab{a}})\citenamefont {Zhang}, \citenamefont {Chen},
  \citenamefont {Yi},\ and\ \citenamefont {Shi}}]{Zhang2025c}%
  \BibitemOpen
  \bibfield  {author} {\bibinfo {author} {\bibfnamefont {W.}~\bibnamefont
  {Zhang}}, \bibinfo {author} {\bibfnamefont {K.}~\bibnamefont {Chen}},
  \bibinfo {author} {\bibfnamefont {S.}~\bibnamefont {Yi}},\ and\ \bibinfo
  {author} {\bibfnamefont {T.}~\bibnamefont {Shi}},\ }\bibfield  {title}
  {\bibinfo {title} {Quantum phases for finite-temperature gases of bosonic
  polar molecules shielded by dual microwaves},\ }\href
  {https://doi.org/10.1103/9cxl-d9zg} {\bibfield  {journal} {\bibinfo
  {journal} {PRX Quantum}\ }\textbf {\bibinfo {volume} {6}},\ \bibinfo {pages}
  {040307} (\bibinfo {year} {2025}{\natexlab{a}})}\BibitemShut {NoStop}%
\bibitem [{\citenamefont {Langen}\ \emph {et~al.}(2025)\citenamefont {Langen},
  \citenamefont {Boronat}, \citenamefont {S\'anchez-Baena}, \citenamefont
  {Bomb\'in}, \citenamefont {Karman},\ and\ \citenamefont
  {Mazzanti}}]{Langen2025a}%
  \BibitemOpen
  \bibfield  {author} {\bibinfo {author} {\bibfnamefont {T.}~\bibnamefont
  {Langen}}, \bibinfo {author} {\bibfnamefont {J.}~\bibnamefont {Boronat}},
  \bibinfo {author} {\bibfnamefont {J.}~\bibnamefont {S\'anchez-Baena}},
  \bibinfo {author} {\bibfnamefont {R.}~\bibnamefont {Bomb\'in}}, \bibinfo
  {author} {\bibfnamefont {T.}~\bibnamefont {Karman}},\ and\ \bibinfo {author}
  {\bibfnamefont {F.}~\bibnamefont {Mazzanti}},\ }\bibfield  {title} {\bibinfo
  {title} {Dipolar droplets of strongly interacting molecules},\ }\href
  {https://doi.org/10.1103/PhysRevLett.134.053001} {\bibfield  {journal}
  {\bibinfo  {journal} {Phys. Rev. Lett.}\ }\textbf {\bibinfo {volume} {134}},\
  \bibinfo {pages} {053001} (\bibinfo {year} {2025})}\BibitemShut {NoStop}%
\bibitem [{\citenamefont {Ciardi}\ \emph {et~al.}(2025)\citenamefont {Ciardi},
  \citenamefont {Pedersen}, \citenamefont {Langen},\ and\ \citenamefont
  {Pohl}}]{Ciardi2025a}%
  \BibitemOpen
  \bibfield  {author} {\bibinfo {author} {\bibfnamefont {M.}~\bibnamefont
  {Ciardi}}, \bibinfo {author} {\bibfnamefont {K.~R.}\ \bibnamefont
  {Pedersen}}, \bibinfo {author} {\bibfnamefont {T.}~\bibnamefont {Langen}},\
  and\ \bibinfo {author} {\bibfnamefont {T.}~\bibnamefont {Pohl}},\ }\bibfield
  {title} {\bibinfo {title} {Self-bound superfluid membranes and monolayer
  crystals of ultracold polar molecules},\ }\href
  {https://doi.org/10.1103/v7gw-xy36} {\bibfield  {journal} {\bibinfo
  {journal} {Phys. Rev. Lett.}\ }\textbf {\bibinfo {volume} {135}},\ \bibinfo
  {pages} {153401} (\bibinfo {year} {2025})}\BibitemShut {NoStop}%
\bibitem [{\citenamefont {Polterauer}\ and\ \citenamefont
  {Zillich}(2025)}]{Polterauer2025a}%
  \BibitemOpen
  \bibfield  {author} {\bibinfo {author} {\bibfnamefont {C.~J.}\ \bibnamefont
  {Polterauer}}\ and\ \bibinfo {author} {\bibfnamefont {R.~E.}\ \bibnamefont
  {Zillich}},\ }\href {https://arxiv.org/abs/2507.18986} {\bibinfo {title}
  {Phases of a {Bose}-{Einstein} condensate of microwave-shielded dipolar
  molecules}} (\bibinfo {year} {2025}),\ \Eprint
  {https://arxiv.org/abs/2507.18986} {arXiv:2507.18986} \BibitemShut {NoStop}%
\bibitem [{\citenamefont {Cardinale}\ \emph {et~al.}(2025)\citenamefont
  {Cardinale}, \citenamefont {Bland},\ and\ \citenamefont
  {Reimann}}]{Cardinale2025a}%
  \BibitemOpen
  \bibfield  {author} {\bibinfo {author} {\bibfnamefont {T.~A.}\ \bibnamefont
  {Cardinale}}, \bibinfo {author} {\bibfnamefont {T.}~\bibnamefont {Bland}},\
  and\ \bibinfo {author} {\bibfnamefont {S.~M.}\ \bibnamefont {Reimann}},\
  }\href {https://arxiv.org/abs/2509.18051} {\bibinfo {title} {Exploring
  molecular supersolidity via exact and mean-field theories: single microwave
  shielding}} (\bibinfo {year} {2025}),\ \Eprint
  {https://arxiv.org/abs/2509.18051} {arXiv:2509.18051} \BibitemShut {NoStop}%
\bibitem [{\citenamefont {Zhang}\ \emph
  {et~al.}(2025{\natexlab{b}})\citenamefont {Zhang}, \citenamefont {Liu},
  \citenamefont {Deng}, \citenamefont {Chen}, \citenamefont {Yi},\ and\
  \citenamefont {Shi}}]{Zhang2025a}%
  \BibitemOpen
  \bibfield  {author} {\bibinfo {author} {\bibfnamefont {W.}~\bibnamefont
  {Zhang}}, \bibinfo {author} {\bibfnamefont {H.}~\bibnamefont {Liu}}, \bibinfo
  {author} {\bibfnamefont {F.}~\bibnamefont {Deng}}, \bibinfo {author}
  {\bibfnamefont {K.}~\bibnamefont {Chen}}, \bibinfo {author} {\bibfnamefont
  {S.}~\bibnamefont {Yi}},\ and\ \bibinfo {author} {\bibfnamefont
  {T.}~\bibnamefont {Shi}},\ }\href {https://arxiv.org/abs/2506.23820}
  {\bibinfo {title} {Supersolid phases in ultracold gases of microwave shielded
  polar molecules}} (\bibinfo {year} {2025}{\natexlab{b}}),\ \Eprint
  {https://arxiv.org/abs/2506.23820} {arXiv:2506.23820} \BibitemShut {NoStop}%
\bibitem [{\citenamefont {Deng}\ \emph {et~al.}(2023)\citenamefont {Deng},
  \citenamefont {Chen}, \citenamefont {Luo}, \citenamefont {Zhang},
  \citenamefont {Yi},\ and\ \citenamefont {Shi}}]{Deng2023a}%
  \BibitemOpen
  \bibfield  {author} {\bibinfo {author} {\bibfnamefont {F.}~\bibnamefont
  {Deng}}, \bibinfo {author} {\bibfnamefont {X.-Y.}\ \bibnamefont {Chen}},
  \bibinfo {author} {\bibfnamefont {X.-Y.}\ \bibnamefont {Luo}}, \bibinfo
  {author} {\bibfnamefont {W.}~\bibnamefont {Zhang}}, \bibinfo {author}
  {\bibfnamefont {S.}~\bibnamefont {Yi}},\ and\ \bibinfo {author}
  {\bibfnamefont {T.}~\bibnamefont {Shi}},\ }\bibfield  {title} {\bibinfo
  {title} {Effective potential and superfluidity of microwave-shielded polar
  molecules},\ }\href {https://doi.org/10.1103/PhysRevLett.130.183001}
  {\bibfield  {journal} {\bibinfo  {journal} {Phys. Rev. Lett.}\ }\textbf
  {\bibinfo {volume} {130}},\ \bibinfo {pages} {183001} (\bibinfo {year}
  {2023})}\BibitemShut {NoStop}%
\bibitem [{\citenamefont {Karman}\ \emph {et~al.}(2025)\citenamefont {Karman},
  \citenamefont {Bigagli}, \citenamefont {Yuan}, \citenamefont {Zhang},
  \citenamefont {Stevenson},\ and\ \citenamefont {Will}}]{Karman2025a}%
  \BibitemOpen
  \bibfield  {author} {\bibinfo {author} {\bibfnamefont {T.}~\bibnamefont
  {Karman}}, \bibinfo {author} {\bibfnamefont {N.}~\bibnamefont {Bigagli}},
  \bibinfo {author} {\bibfnamefont {W.}~\bibnamefont {Yuan}}, \bibinfo {author}
  {\bibfnamefont {S.}~\bibnamefont {Zhang}}, \bibinfo {author} {\bibfnamefont
  {I.}~\bibnamefont {Stevenson}},\ and\ \bibinfo {author} {\bibfnamefont
  {S.}~\bibnamefont {Will}},\ }\bibfield  {title} {\bibinfo {title} {Double
  microwave shielding},\ }\href {https://doi.org/10.1103/b8pm-3prn} {\bibfield
  {journal} {\bibinfo  {journal} {PRX Quantum}\ }\textbf {\bibinfo {volume}
  {6}},\ \bibinfo {pages} {020358} (\bibinfo {year} {2025})}\BibitemShut
  {NoStop}%
\bibitem [{\citenamefont {Deng}\ \emph {et~al.}(2025)\citenamefont {Deng},
  \citenamefont {Hu}, \citenamefont {Jin}, \citenamefont {Yi},\ and\
  \citenamefont {Shi}}]{Deng2025a}%
  \BibitemOpen
  \bibfield  {author} {\bibinfo {author} {\bibfnamefont {F.}~\bibnamefont
  {Deng}}, \bibinfo {author} {\bibfnamefont {X.}~\bibnamefont {Hu}}, \bibinfo
  {author} {\bibfnamefont {W.-J.}\ \bibnamefont {Jin}}, \bibinfo {author}
  {\bibfnamefont {S.}~\bibnamefont {Yi}},\ and\ \bibinfo {author}
  {\bibfnamefont {T.}~\bibnamefont {Shi}},\ }\bibfield  {title} {\bibinfo
  {title} {Two- and many-body physics of ultracold molecules dressed by dual
  microwave fields},\ }\href {https://doi.org/10.1038/s41467-025-66067-2}
  {\bibfield  {journal} {\bibinfo  {journal} {Nat. Commun.}\ }\textbf {\bibinfo
  {volume} {16}},\ \bibinfo {pages} {11219} (\bibinfo {year}
  {2025})}\BibitemShut {NoStop}%
\bibitem [{\citenamefont {Schindewolf}\ \emph {et~al.}(2026)\citenamefont
  {Schindewolf}, \citenamefont {Hertkorn}, \citenamefont {Stevenson},
  \citenamefont {Ciardi}, \citenamefont {Gro{\ss}}, \citenamefont {Wang},
  \citenamefont {Karman}, \citenamefont {Qu{\'e}m{\'e}ner}, \citenamefont
  {Will}, \citenamefont {Pohl},\ and\ \citenamefont
  {Langen}}]{Schindewolf2026a}%
  \BibitemOpen
  \bibfield  {author} {\bibinfo {author} {\bibfnamefont {A.}~\bibnamefont
  {Schindewolf}}, \bibinfo {author} {\bibfnamefont {J.}~\bibnamefont
  {Hertkorn}}, \bibinfo {author} {\bibfnamefont {I.}~\bibnamefont {Stevenson}},
  \bibinfo {author} {\bibfnamefont {M.}~\bibnamefont {Ciardi}}, \bibinfo
  {author} {\bibfnamefont {P.}~\bibnamefont {Gro{\ss}}}, \bibinfo {author}
  {\bibfnamefont {D.}~\bibnamefont {Wang}}, \bibinfo {author} {\bibfnamefont
  {T.}~\bibnamefont {Karman}}, \bibinfo {author} {\bibfnamefont
  {G.}~\bibnamefont {Qu{\'e}m{\'e}ner}}, \bibinfo {author} {\bibfnamefont
  {S.}~\bibnamefont {Will}}, \bibinfo {author} {\bibfnamefont {T.}~\bibnamefont
  {Pohl}},\ and\ \bibinfo {author} {\bibfnamefont {T.}~\bibnamefont {Langen}},\
  }\bibfield  {title} {\bibinfo {title} {Colloquium: Strongly dipolar molecular
  bose-einstein condensates: From few- to many-body physics},\ }\bibfield
  {journal} {\bibinfo  {journal} {Rev. Mod. Phys.}\ }\href
  {https://doi.org/10.1103/r55l-f93m} {10.1103/r55l-f93m} (\bibinfo {year}
  {2026})\BibitemShut {NoStop}%
\bibitem [{Note1()}]{Note1}%
  \BibitemOpen
  \bibinfo {note} {Using $a_{\protect \mathrm {d}m}$ defined in \cite
  {Zhang2026a}, our $\epsilon _m = a_{\protect \mathrm {d}m}/3a_s$ for $m=0$,
  $2$. Using $\epsilon _{3,2}$ defined in \cite {Zhang2025a}, our $\epsilon
  _2=\epsilon _{3,2}/\protect \sqrt 3$.}\BibitemShut {Stop}%
\bibitem [{Note2()}]{Note2}%
  \BibitemOpen
  \bibinfo {note} {Not to be confused with the vacuum
  permittivity.}\BibitemShut {Stop}%
\bibitem [{\citenamefont {Yuan}\ \emph {et~al.}(2025)\citenamefont {Yuan},
  \citenamefont {Zhang}, \citenamefont {Bigagli}, \citenamefont {Kwak},
  \citenamefont {Warner}, \citenamefont {Karman}, \citenamefont {Stevenson},\
  and\ \citenamefont {Will}}]{Yuan2025a}%
  \BibitemOpen
  \bibfield  {author} {\bibinfo {author} {\bibfnamefont {W.}~\bibnamefont
  {Yuan}}, \bibinfo {author} {\bibfnamefont {S.}~\bibnamefont {Zhang}},
  \bibinfo {author} {\bibfnamefont {N.}~\bibnamefont {Bigagli}}, \bibinfo
  {author} {\bibfnamefont {H.}~\bibnamefont {Kwak}}, \bibinfo {author}
  {\bibfnamefont {C.}~\bibnamefont {Warner}}, \bibinfo {author} {\bibfnamefont
  {T.}~\bibnamefont {Karman}}, \bibinfo {author} {\bibfnamefont
  {I.}~\bibnamefont {Stevenson}},\ and\ \bibinfo {author} {\bibfnamefont
  {S.}~\bibnamefont {Will}},\ }\href {https://arxiv.org/abs/2505.08773}
  {\bibinfo {title} {Extreme loss suppression and wide tunability of dipolar
  interactions in an ultracold molecular gas}} (\bibinfo {year} {2025}),\
  \Eprint {https://arxiv.org/abs/2505.08773} {arXiv:2505.08773} \BibitemShut
  {NoStop}%
\bibitem [{\citenamefont {Lima}\ and\ \citenamefont
  {Pelster}(2011)}]{Lima2011a}%
  \BibitemOpen
  \bibfield  {author} {\bibinfo {author} {\bibfnamefont {A.~R.~P.}\
  \bibnamefont {Lima}}\ and\ \bibinfo {author} {\bibfnamefont {A.}~\bibnamefont
  {Pelster}},\ }\bibfield  {title} {\bibinfo {title} {Quantum fluctuations in
  dipolar {Bose} gases},\ }\href {https://doi.org/10.1103/PhysRevA.84.041604}
  {\bibfield  {journal} {\bibinfo  {journal} {Phys. Rev. A}\ }\textbf {\bibinfo
  {volume} {84}},\ \bibinfo {pages} {041604(R)} (\bibinfo {year}
  {2011})}\BibitemShut {NoStop}%
\bibitem [{\citenamefont {Lima}\ and\ \citenamefont
  {Pelster}(2012)}]{Lima2012a}%
  \BibitemOpen
  \bibfield  {author} {\bibinfo {author} {\bibfnamefont {A.~R.~P.}\
  \bibnamefont {Lima}}\ and\ \bibinfo {author} {\bibfnamefont {A.}~\bibnamefont
  {Pelster}},\ }\bibfield  {title} {\bibinfo {title} {Beyond mean-field
  low-lying excitations of dipolar {Bose} gases},\ }\href
  {https://doi.org/10.1103/PhysRevA.86.063609} {\bibfield  {journal} {\bibinfo
  {journal} {Phys. Rev. A}\ }\textbf {\bibinfo {volume} {86}},\ \bibinfo
  {pages} {063609} (\bibinfo {year} {2012})}\BibitemShut {NoStop}%
\bibitem [{Note3()}]{Note3}%
  \BibitemOpen
  \bibinfo {note} {By taking the Fourier transform, changing variables to
  $u_i=l_ik_i$, integrating the angular variables, and changing variables back
  to $k_i$.}\BibitemShut {Stop}%
\bibitem [{\citenamefont {Giovanazzi}\ \emph {et~al.}(2006)\citenamefont
  {Giovanazzi}, \citenamefont {Pedri}, \citenamefont {Santos}, \citenamefont
  {Griesmaier}, \citenamefont {Fattori}, \citenamefont {Koch}, \citenamefont
  {Stuhler},\ and\ \citenamefont {Pfau}}]{Giovanazzi2006a}%
  \BibitemOpen
  \bibfield  {author} {\bibinfo {author} {\bibfnamefont {S.}~\bibnamefont
  {Giovanazzi}}, \bibinfo {author} {\bibfnamefont {P.}~\bibnamefont {Pedri}},
  \bibinfo {author} {\bibfnamefont {L.}~\bibnamefont {Santos}}, \bibinfo
  {author} {\bibfnamefont {A.}~\bibnamefont {Griesmaier}}, \bibinfo {author}
  {\bibfnamefont {M.}~\bibnamefont {Fattori}}, \bibinfo {author} {\bibfnamefont
  {T.}~\bibnamefont {Koch}}, \bibinfo {author} {\bibfnamefont {J.}~\bibnamefont
  {Stuhler}},\ and\ \bibinfo {author} {\bibfnamefont {T.}~\bibnamefont
  {Pfau}},\ }\bibfield  {title} {\bibinfo {title} {Expansion dynamics of a
  dipolar {Bose}-{Einstein} condensate},\ }\href
  {https://doi.org/10.1103/PhysRevA.74.013621} {\bibfield  {journal} {\bibinfo
  {journal} {Phys. Rev. A}\ }\textbf {\bibinfo {volume} {74}},\ \bibinfo
  {pages} {013621} (\bibinfo {year} {2006})}\BibitemShut {NoStop}%
\bibitem [{\citenamefont {Glaum}\ and\ \citenamefont
  {Pelster}(2007)}]{Glaum2007b}%
  \BibitemOpen
  \bibfield  {author} {\bibinfo {author} {\bibfnamefont {K.}~\bibnamefont
  {Glaum}}\ and\ \bibinfo {author} {\bibfnamefont {A.}~\bibnamefont
  {Pelster}},\ }\bibfield  {title} {\bibinfo {title} {{Bose}-{Einstein}
  condensation temperature of dipolar gas in anisotropic harmonic trap},\
  }\href {https://doi.org/10.1103/PhysRevA.76.023604} {\bibfield  {journal}
  {\bibinfo  {journal} {Phys. Rev. A}\ }\textbf {\bibinfo {volume} {76}},\
  \bibinfo {pages} {023604} (\bibinfo {year} {2007})}\BibitemShut {NoStop}%
\bibitem [{Note4()}]{Note4}%
  \BibitemOpen
  \bibinfo {note} {We align our dipoles along $z$ whereas \cite
  {Giovanazzi2006a} aligns along $x$, but our function $f(x,y)$ is exactly the
  same as theirs.}\BibitemShut {Stop}%
\bibitem [{Note5()}]{Note5}%
  \BibitemOpen
  \bibinfo {note} {For $0<y\ll 1$, as $x$ is reduced $f_2(x,y)$ changes sharply
  from $\approx -1$ for $x\gg y$, through zero at $x=y$, to $1$ at
  $x=0$.}\BibitemShut {Stop}%
\bibitem [{\citenamefont {Santos}\ \emph {et~al.}(2000)\citenamefont {Santos},
  \citenamefont {Shlyapnikov}, \citenamefont {Zoller},\ and\ \citenamefont
  {Lewenstein}}]{Santos2000a}%
  \BibitemOpen
  \bibfield  {author} {\bibinfo {author} {\bibfnamefont {L.}~\bibnamefont
  {Santos}}, \bibinfo {author} {\bibfnamefont {G.~V.}\ \bibnamefont
  {Shlyapnikov}}, \bibinfo {author} {\bibfnamefont {P.}~\bibnamefont
  {Zoller}},\ and\ \bibinfo {author} {\bibfnamefont {M.}~\bibnamefont
  {Lewenstein}},\ }\bibfield  {title} {\bibinfo {title} {{Bose}-{Einstein}
  condensation in trapped dipolar gases},\ }\href
  {https://doi.org/10.1103/PhysRevLett.85.1791} {\bibfield  {journal} {\bibinfo
   {journal} {Phys. Rev. Lett.}\ }\textbf {\bibinfo {volume} {85}},\ \bibinfo
  {pages} {1791} (\bibinfo {year} {2000})}\BibitemShut {NoStop}%
\bibitem [{\citenamefont {Ferrier-Barbut}\ \emph {et~al.}(2016)\citenamefont
  {Ferrier-Barbut}, \citenamefont {Kadau}, \citenamefont {Schmitt},
  \citenamefont {Wenzel},\ and\ \citenamefont {Pfau}}]{Ferrier-Barbut2016a}%
  \BibitemOpen
  \bibfield  {author} {\bibinfo {author} {\bibfnamefont {I.}~\bibnamefont
  {Ferrier-Barbut}}, \bibinfo {author} {\bibfnamefont {H.}~\bibnamefont
  {Kadau}}, \bibinfo {author} {\bibfnamefont {M.}~\bibnamefont {Schmitt}},
  \bibinfo {author} {\bibfnamefont {M.}~\bibnamefont {Wenzel}},\ and\ \bibinfo
  {author} {\bibfnamefont {T.}~\bibnamefont {Pfau}},\ }\bibfield  {title}
  {\bibinfo {title} {Observation of quantum droplets in a strongly dipolar
  {Bose} gas},\ }\href {https://doi.org/10.1103/PhysRevLett.116.215301}
  {\bibfield  {journal} {\bibinfo  {journal} {Phys. Rev. Lett.}\ }\textbf
  {\bibinfo {volume} {116}},\ \bibinfo {pages} {215301} (\bibinfo {year}
  {2016})}\BibitemShut {NoStop}%
\bibitem [{\citenamefont {Dalibard}(2024)}]{Dalibard2024a}%
  \BibitemOpen
  \bibfield  {author} {\bibinfo {author} {\bibfnamefont {J.}~\bibnamefont
  {Dalibard}},\ }\href
  {https://pro.college-de-france.fr/jean.dalibard/CdF/2024/total_en_2024.pdf}
  {\bibinfo {title} {Magnetic interactions between cold atoms: Quantum droplets
  and supersolid states}},\ \bibinfo {howpublished} {Coll{\`e}ge de France,
  Lecture Series} (\bibinfo {year} {2024})\BibitemShut {NoStop}%
\bibitem [{\citenamefont {Baillie}\ \emph {et~al.}(2017)\citenamefont
  {Baillie}, \citenamefont {Wilson},\ and\ \citenamefont
  {Blakie}}]{Baillie2017a}%
  \BibitemOpen
  \bibfield  {author} {\bibinfo {author} {\bibfnamefont {D.}~\bibnamefont
  {Baillie}}, \bibinfo {author} {\bibfnamefont {R.~M.}\ \bibnamefont
  {Wilson}},\ and\ \bibinfo {author} {\bibfnamefont {P.~B.}\ \bibnamefont
  {Blakie}},\ }\bibfield  {title} {\bibinfo {title} {Collective excitations of
  self-bound droplets of a dipolar quantum fluid},\ }\href
  {https://doi.org/10.1103/PhysRevLett.119.255302} {\bibfield  {journal}
  {\bibinfo  {journal} {Phys. Rev. Lett.}\ }\textbf {\bibinfo {volume} {119}},\
  \bibinfo {pages} {255302} (\bibinfo {year} {2017})}\BibitemShut {NoStop}%
\bibitem [{\citenamefont {Baillie}(2026)}]{Baillie2026b}%
  \BibitemOpen
  \bibfield  {author} {\bibinfo {author} {\bibfnamefont {D.}~\bibnamefont
  {Baillie}},\ }\bibfield  {title} {\bibinfo {title} {Symmetry and self-bound
  droplets in dipolar molecular gases [data set]},\ }\href
  {https://doi.org/10.5281/zenodo.20078499} {10.5281/zenodo.20078499} (\bibinfo
  {year} {2026})\BibitemShut {NoStop}%
\bibitem [{\citenamefont {Jackson}(1999)}]{JacksonBook1999a}%
  \BibitemOpen
  \bibfield  {author} {\bibinfo {author} {\bibfnamefont {J.~D.}\ \bibnamefont
  {Jackson}},\ }\href@noop {} {\emph {\bibinfo {title} {Classical
  electrodynamics}}},\ \bibinfo {edition} {3rd}\ ed.\ (\bibinfo  {publisher}
  {Wiley},\ \bibinfo {address} {New York},\ \bibinfo {year} {1999})\BibitemShut
  {NoStop}%
\bibitem [{\citenamefont {Ronen}\ \emph {et~al.}(2006)\citenamefont {Ronen},
  \citenamefont {Bortolotti},\ and\ \citenamefont {Bohn}}]{Ronen2006a}%
  \BibitemOpen
  \bibfield  {author} {\bibinfo {author} {\bibfnamefont {S.}~\bibnamefont
  {Ronen}}, \bibinfo {author} {\bibfnamefont {D.~C.~E.}\ \bibnamefont
  {Bortolotti}},\ and\ \bibinfo {author} {\bibfnamefont {J.~L.}\ \bibnamefont
  {Bohn}},\ }\bibfield  {title} {\bibinfo {title} {{Bogoliubov} modes of a
  dipolar condensate in a cylindrical trap},\ }\href
  {https://doi.org/10.1103/PhysRevA.74.013623} {\bibfield  {journal} {\bibinfo
  {journal} {Phys. Rev. A}\ }\textbf {\bibinfo {volume} {74}},\ \bibinfo
  {pages} {013623} (\bibinfo {year} {2006})}\BibitemShut {NoStop}%
\bibitem [{\citenamefont {Saito}(2016)}]{Saito2016a}%
  \BibitemOpen
  \bibfield  {author} {\bibinfo {author} {\bibfnamefont {H.}~\bibnamefont
  {Saito}},\ }\bibfield  {title} {\bibinfo {title} {Path-integral {Monte}
  {Carlo} study on a droplet of a dipolar {Bose}-{Einstein} condensate
  stabilized by quantum fluctuation},\ }\href
  {https://doi.org/10.7566/JPSJ.85.053001} {\bibfield  {journal} {\bibinfo
  {journal} {J. Phys. Soc. Jpn}\ }\textbf {\bibinfo {volume} {85}},\ \bibinfo
  {pages} {053001} (\bibinfo {year} {2016})}\BibitemShut {NoStop}%
\bibitem [{Note6()}]{Note6}%
  \BibitemOpen
  \bibinfo {note} {The prefactor of Eq.~(A10) of \cite {Bisset2016a} should be
  $1/128$, not $1/64$.}\BibitemShut {Stop}%
\bibitem [{Note7()}]{Note7}%
  \BibitemOpen
  \bibinfo {note} {For larger parameters $\protect \mathcal {Q}_5$ increases
  like $|\epsilon _0|^{5/2}$ or $|\epsilon _2|^{5/2}$ [see Eq.~\protect \eqref
  {e:Q5}], but the imaginary part of $\protect \mathcal {Q}_5$ will then be
  significant.}\BibitemShut {Stop}%
\end{thebibliography}


\end{document}